\documentclass[12pt,a4paper,aps,prd,preprint,superscriptaddress,nofootinbib]{revtex4-1}
\usepackage[utf8]{inputenc}
\usepackage{graphicx}
\usepackage{amssymb}
\usepackage{textcomp}
\usepackage{amsmath}
\usepackage{tabularx}
\usepackage{bm}
\usepackage{times}
\usepackage{color}
\usepackage{slashed}
\usepackage{multirow}
\usepackage{verbatim}
\usepackage{cancel}
\usepackage{subfigure}
\usepackage[normalem]{ulem}
\usepackage{float}

\usepackage[colorlinks=true, pdfstartview=FitV, linkcolor=blue, citecolor=blue, urlcolor=blue]{hyperref}
\allowdisplaybreaks[4]

\def\lsim{\mathrel{\raise.3ex\hbox{$<$\kern-.75em\lower1ex\hbox{$\sim$}}}}
\def\gsim{\mathrel{\raise.3ex\hbox{$>$\kern-.75em\lower1ex\hbox{$\sim$}}}}

\definecolor{orange}{rgb}{1,0.5,0}

\begin{document}

\title{Solutions to axion electrodynamics with electric-magnetic duality in supersymmetric Seiberg-Witten theory}

\author{Tong Li}
\email{litong@nankai.edu.cn}
\affiliation{
School of Physics, Nankai University, Tianjin 300071, China
}
\author{Rui-Jia Zhang}
\email{zhangruijia@mail.nankai.edu.cn}
\affiliation{
School of Physics, Nankai University, Tianjin 300071, China
}

\begin{abstract}
Axion and magnetic monopole are among the most fascinating candidates for physics
beyond the Standard Model. The potential connection between axion and magnetic monopole stems from the Witten effect and is revealed by non-standard axion electrodynamics. Non-standard axion electrodynamics under electric-magnetic duality modifies conventional axion Maxwell equations and motivates intriguing axion-photon phenomenology. A calculable ultraviolet model of Peccei-Quinn axion coupled to magnetic monopoles and electric charges was proposed based on $\mathcal{N}=2$ supersymmetric Seiberg-Witten (SW) theory with manifest electric-magnetic duality. In this work, we aim to investigate the solutions to the non-linear axion electrodynamics from SW axion model and propose relevant detection strategies for non-trivial axion-photon couplings. Based on the infrared Lagrangian of SW axion, we derive the electromagetic (EM) equations of motion.
We also analyze the moduli space coordinate in SW theory and find out the reliabe parameter space. We then solve the resultant axion Maxwell equations with an external EM field. The observable axion-induced EM fields are obtained analytically and then numerically computed. Finally, we propose the detection strategy with an LC
circuit and show the prospective sensitivity to SW axion-photon couplings.
\end{abstract}

\maketitle

\tableofcontents

%%%%%%%%%%%%%%%%%%%%%%%%%%%%%%%%
\section{Introduction}
\label{sec:Intro}
%%%%%%%%%%%%%%%%%%%%%%%%%%%%%%%%

Axion and magnetic monopole are two of the most interesting and mysterious candidates of physics
beyond the Standard Model (SM). Axion is a hypothetical pseudo-scalar particle arised as a solution to the strong CP problem in quantum chromodynamics (QCD). The Peccei-Quinn (PQ) mechanism introduces a pseudo-Goldstone boson $a$ after the spontaneous breaking of a QCD anomalous global $U(1)_{\rm PQ}$ symmetry~\cite{Peccei:1977hh,Peccei:1977ur,Weinberg:1977ma,Wilczek:1977pj,Baluni:1978rf,Crewther:1979pi,Kim:1979if,Shifman:1979if,Dine:1981rt,Zhitnitsky:1980tq,Baker:2006ts,Pendlebury:2015lrz}. Both the QCD axion~\cite{Peccei:1977hh,Peccei:1977ur,Weinberg:1977ma,Wilczek:1977pj} (see a recent review Ref.~\cite{DiLuzio:2020wdo}) and axion-like particle (ALP)~\cite{Kim:1986ax,Kuster:2008zz} can serve as dark matter (DM) through the misalignment mechanism~\cite{Preskill:1982cy,Dine:1982ah}. The existence of magnetic charges was initially motivated by the consideration
of electric-magnetic duality in classical electromagnetism. In 1931, Dirac suggested the existence of magnetic monopole in quantum theory~\cite{Dirac:1931kp}. The Dirac monopole can arise from QCD~\cite{Wu:1975es} to the grand unification theory~\cite{tHooft:1974kcl,Polyakov:1974ek}. The search for axions and magnetic monopoles also stands at the forefront of the experimental particle physics.

In 1979, E. Witten first realized the so-called Witten effect which implied the possible connection between axion and magnetic monopole. He pointed out that a non-zero vacuum angle $\theta$ in the CP violating term $\theta F^{\mu\nu}\tilde{F}_{\mu\nu}$ introduces an electric charge $-\theta e/2\pi$ for magnetic monopoles~\cite{Witten:1979ey}.
Later, W. Fischler et al. derived the consequent axion-dyon dynamics under the classical electromagnetism~\cite{Fischler:1983sc}. Recently, Refs.~\cite{Sokolov:2022fvs,Sokolov:2023pos} considered the electric-magnetic duality with axion coupled to magnetically charged particle under the quantum electromagnetodynamics (QEMD) built by Schwinger and Zwanziger~\cite{Schwinger:1966nj,Zwanziger:1968rs,Zwanziger:1970hk}. They constructed a non-standard axion electrodynamics in the low-energy effective
field theory (EFT) with distinct form of axion-photon couplings. As a result, this non-standard axion electrodynamics modifies conventional axion Maxwell equations~\cite{Sokolov:2022fvs,Li:2022oel,Tobar:2022rko} and motivates interesting axion phenomenology~\cite{Sokolov:2022fvs,Li:2022oel,Tobar:2022rko,McAllister:2022ibe,Li:2023kfh,Li:2023aow,Tobar:2023rga,Patkos:2023lof,Dai:2024dkr,Dai:2024zim}.

However, Ref.~\cite{Heidenreich:2023pbi} showed that the non-standard axion electrodynamics violates the usual quantization rule of axion-photon coupling and is not phenomenologically viable. This realization motivates theorists to reconsider reliable axion model with consistent electric-magnetic duality and coupling quantization. More recently, Ref.~\cite{Csaki:2024plt} introduced a calculable ultraviolet (UV) model of PQ axion coupled to magnetic monopoles and electric charges. This axion model is based on $\mathcal{N}=2$ supersymmetric Seiberg-Witten (SW) theory~\cite{Seiberg:1994rs,Seiberg:1994aj} with manifest electric-magnetic duality. This SW theory has global $SU(2)_R\times U(1)_\mathcal{R}$ symmetry and the set of all vacua in it refers to a moduli space.
After the breaking of $SU(2)$ to $U(1)$, the low-energy effective action can be described by $\mathcal{N}=1$ chiral superfield $A$ in electric frame (E-frame) and its dual field $A_D$ in magnetic frame (M-frame). They also undergo non-trivial $SL(2,\mathbb{Z})$ transformations and are related to each other through their ``prepotential'' $\mathcal{F}(A)$ or $\mathcal{F}_D(A_D)$. The holomorphic gauge couplings $\tau$ and $\tau_D$ on the moduli space are respectively determined by prepotential $\mathcal{F}(A)$ and $\mathcal{F}_D(A_D)$. The perturbative parts of the prepotential $\mathcal{F}(A)$ and its dual $\mathcal{F}_D(A_D)$ arise from one-loop
contributions of massive gauge multiplets and light monopoles, respectively. The series
terms in the prepotentials attribute to non-perturbative instanton effects. The $SL(2,\mathbb{Z})$ S-duality provides a ``window'' through
which the originally non-perturbative, strongly coupled physics on the moduli space can be mapped onto perturbative descriptions.

The spontaneous breaking of $U(1)_\mathcal{R}$ plays a role analogous to that of the PQ symmetry. The analogous breaking
of the $U(1)_\mathcal{R}$ symmetry on the moduli space gives rise to the axion field $a$ as a phase of $A$ (or dual axion $a_D$ as a phase of $A_D$) in this model. The non-perturbative instanton effects in the prepotential induce non-linear axion-photon couplings. Besides usual anomalous axion-photon coupling as linear term, instanton effect introduces non-perturbative corrections
due to the monopoles in the E-frame. By contrast, in the M-frame, the magnetic charges introduce linear coupling of the dual axion $a_D$ and the electric charges contribute to non-perturbative effects.

In this work, we aim to investigate the solutions to the non-linear axion electrodynamics from SW theory and propose relevant detection strategies. In both E-frame and M-frame, based on the infrared (IR) Lagrangian of SW axion, we derive the electromagetic (EM) equations of motion (EoMs). Then, the expansion of $\tau$ or $\tau_D$ is used to simplify the EoMs.
We also analyze the parameter space of moduli space coordinate in both frames and figure out the reliabe parameter region. We then solve the resultant axion Maxwell equations under two cases with only an external magnetic field or only an external electric field. The observable axion-induced EM fields are obtained and numerically computed. Finally, we propose the detection setup with an LC
circuit~\cite{Sikivie:2013laa,Irastorza:2018dyq,Sikivie:2020zpn} to probe the axion-photon couplings and show the prospective sensitivity in these two frames.

This paper is organized as follows. In Sec.~\ref{sec:SW}, we review $\mathcal{N}=2$ supersymmetric Seiberg-Witten theory and the SW axion model with electric-magnetic duality. The solutions to SW axion electrodynamics are derived in Sec.~\ref{sec:Solution}. We also analyze the parameter space of moduli space coordinate in both E-frame and M-frame. We then propose experimental strategies and show the prospective detection results of SW axion-photon couplings in Sec.~\ref{sec:Sensitivity}. Our conclusions are drawn in Sec.~\ref{sec:Con}.

%%%%%%%%%%%%%%%%%%%%%%%%%%%%%%%%%%%
\section{Seiberg-Witten theory and axion model with electric-magnetic duality}
\label{sec:SW}
%%%%%%%%%%%%%%%%%%%%%%%%%%%%%%%%%%%
%%%%%%%%%%%%%%%%%%%%%%
\subsection{$\mathcal{N}=2$ supersymmetric Seiberg-Witten theory}
%%%%%%%%%%%%%%%%%%%%%%

\begin{figure}[htbp!]
\centering
\includegraphics[width=0.6\linewidth]{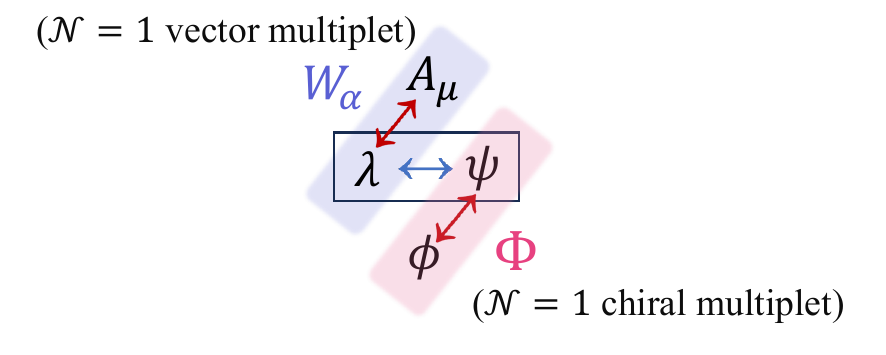}
\caption{The field structure for $\mathcal{N}=2$ supersymmetric $SU(2)$ gauge theory. The red arrows represent the $\mathcal{N}=1$ sub-supersymmetry generator manifest in the $\mathcal{N}=1$
superfield formalism. The blue arrow indicates a global symmetry $SU(2)_R$ relating the fermions $\lambda,\psi$ within the $\mathcal{N}=2$ multiplet.}
\label{fig:multiplet}
\end{figure}

The Seiberg-Witten (SW) theory in this work refers to the low-energy effective action of $\mathcal{N}=2$ supersymmetric $SU(2)$ gauge theory~\cite{Seiberg:1994rs,Seiberg:1994aj} (see Ref.~\cite{Tachikawa:2013kta} for a recent lecture). We first review the $\mathcal{N}=2$ supersymmetric SW theory. In $\mathcal{N}=2$ supersymmetric theories having global $SU(2)_R\times U(1)_{\mathcal{R}}$ symmetry, the most fundamental supermultiplet is shown in Fig.~\ref{fig:multiplet}.
This multiplet consists of $\mathcal{N}=1$ vector multiplet $W_\alpha$ and $\mathcal{N}=1$ chiral multiplet $\Phi$.
The vector multiplet $W_\alpha$ contains a gauge vector field $A_\mu$ and a Weyl fermion $\lambda$. The chiral multiplet $\Phi$ contains a complex scalar field $\phi$ and another Weyl fermion $\psi$. $A_\mu$ and $\phi$ are singlets under $SU(2)_R$ symmetry and $\lambda,\psi$ forms a doublet. The chiral multiplet $\Phi$ has R-charge 2 under the other $U(1)_\mathcal{R}$ symmetry.

In this theory, different vacuum states can be parameterized by an $SU(2)$ gauge invariant quantity $u\equiv\frac{1}{2}\text{tr}(\phi^2)$, and the set of all such vacua is referred to a moduli space.
The parameter $u$ can be regarded as a coordinate on this continuous moduli space.
When the scalar field $\phi$ acquires a non-zero vacuum expectation value (vev) on this moduli space, the gauge symmetry breaking reduces $SU(2)$ to $U(1)$. Then, only the gauge bosons $A^1_\mu$ and $A^2_\mu$ acqurie mass by absorbing the Goldstone degrees of freedom from $\phi$. The remaining gauge boson $A^3_\mu$ stays massless, referred to a photon. It implies that, on this moduli space, the SW theory retains a massless Abelian superfield component after gauge symmetry breaking. This remaining degree of freedom can be analogized to electromagnetism in the low-energy theory. Therefore, one calls this branch of the moduli space with continuous vacua as the ``Coulomb branch''.

After the breaking of $SU(2)$ to $U(1)$, we are ready to work in the IR regime with the photon-like field in the low-energy electromagnetic dynamics
\begin{eqnarray}
SU(2):~(\phi,A_\mu)\to U(1):~(A,V_\mu)\;, 
\end{eqnarray}
where $A$ ($V_\mu$) denotes a $\mathcal{N}=1$ chiral (vector) superfield.
Similarly, we can also define the dual form of the vector mulitiplet in IR theory, i.e., the ``dual photon'' denoted by $(A_D,V_D)$.~\footnote{Here we adopt the convention for chiral superfield in Ref.~\cite{Csaki:2024plt} in order to avoid confusion with the axion field~\cite{Tachikawa:2013kta}.} As the two mutually dual complex scalars preserved in the SW theory, the scalar components $A$ and $A_D$ are holomorphic functions of $u$ and thus contain properties such as monodromy transformation. When $A$ or $A_D$ traverses a closed loop around a singularity of $u$ on the moduli space, their values undergo a change.
This corresponds to performing an electromagnetic duality transformation under $SL(2,\mathbb{Z})$.
Mathematically, the chiral superfield of the IR theory is entirely governed by holomorphic functions $A$ and $A_D$.
As a result, the low-energy effective action can be fully encoded by the integrals of holomorphic functions. They are known as the ``prepotential'' $\mathcal{F}(A)$, and the dual prepotential $\mathcal{F}_D(A_D)$. $\mathcal{F}_D(A_D)$ is related to $\mathcal{F}(A)$ by a Legendre transformation
\begin{eqnarray}
    A_D=\frac{\partial \mathcal{F}(A)}{\partial A}\;,A=-\frac{\partial\mathcal{F}_D(A_D)}{\partial A_D}\;.
    \label{eq:defAAD}
\end{eqnarray}
The $SU(2)$ Seiberg-Witten solution can be obtained by constructing an elliptic curve~\cite{Tachikawa:2013kta}
\begin{eqnarray}
    \Sigma:\Lambda^2z+\frac{\Lambda^2}{z}=x^2-u\;,
    \label{equ:epcurve}
\end{eqnarray}
where $\Lambda$ is the dynamical scale of the theory, $u$ is the moduli space coordinate introduced above, and $x,z$ are both quantities on the complex plane. One then defines a differential form $\lambda=xdz/z$, such that $A$ and $A_D$ can be expressed as two one-dimensional intergals of this differential along the two fundamental cycles $C_1,C_2$ on the curve
\begin{eqnarray}
    A(u)=\frac{1}{2\pi i}\oint_{C_1}\lambda\;,~A_D(u)=\frac{1}{2\pi i}\oint_{C_2}\lambda\;.
\end{eqnarray}
Then, the vevs of these complex scalar components in SW theory can be given by the Gauss hypergeometric function on the moduli space. They are expressed as explicit functional forms of the moduli coordinate $u$~\cite{Csaki:2024plt}
\begin{eqnarray}
    A^v(u)&=&\sqrt{u+2\Lambda^2} _2F_1\left(-\frac{1}{2},\frac{1}{2},1;\frac{4\Lambda^2}{u+2\Lambda^2}\right)\;,
    \label{eq:AAD1}\\
    A^v_D(u)&=&i\frac{u-2\Lambda^2}{2\Lambda}~_2F_1\left(\frac{1}{2},\frac{1}{2},2;\frac{2\Lambda^2-u}{4\Lambda^2}\right)\;,
    \label{eq:AAD2}
\end{eqnarray}
where the superscript ``$v$'' is kept in order to distinguish them from the scalar field components $A(u)$ and $A_D(u)$.

It is known that the hypergeometric function has multiple branch points, which leads to the existence of branch points in $A(u)$ and $A_D(u)$. The origin of this phenomenon stems from the constructed elliptic curve Eq.~\eqref{equ:epcurve}. It is a two-sheeted surface with several branch points: $z = 0, \infty$, and $z_\pm = \frac{\omega \pm \sqrt{\omega^2-4}}{2}$ with $\omega = z + \frac{1}{z}$. 
In general, the $z_\pm$ branch points geometrically correspond to two points on the upper sheet of the SW curve and the other two on the lower sheet. In particular, when $\omega^2 = 4$, i.e., $z_\pm=\pm1\to u = \pm 2\Lambda^2$, the discriminant of the quadratic equation vanishes and then these four branch degenerate into two. Consequently, for the SW theory, the mathematically independent branch points are only $z = 0, \infty$ and $z_\pm=\pm1~(u=\pm 2\Lambda^2)$.

The points $z = 0$ and $\infty$ correspond to the coordinate at infinity $u = \infty$ on the moduli space. This coordinate represents the weak coupling (perturbative) region of the theory where it reduces to a free $\mathcal{N}=2$ $U(1)$ gauge theory. In contrast, the vicinity of $u = \pm 2\Lambda^2$ corresponds to the strongly coupled region of the theory. Obviously, $u = \pm 2\Lambda^2$ are the two points where non-perturbative physics takes place. At these points, specific Bogomol'nyi-Prasad-Sommerfield (BPS) states become massless.
At $u=+2\Lambda^2$, the massless magnetic monopole with $({\rm magnetic~charge}\equiv g,{\rm electric~charge}\equiv q)=(1,0)$ emerges, while at $u=-2\Lambda^2$ the dyon with charge $(g,q)=(1,2)$ becomes massless.
More importantly, when $u$ encircles either of these branch points once, $A(u)$ and $A_D(u)$ undergo non-trivial $SL(2,\mathbb{Z})$ transformations. They correspond to S-duality transformation and R-shift transformation, respectively.

In $\mathcal{N}=2$ supersymmetric gauge theory, the holomorphic gauge coupling $\tau$ is defined as
\begin{eqnarray}
\tau=\frac{\partial^2\mathcal{F}}{\partial A^2}\equiv \frac{\theta}{2\pi}+\frac{4\pi i}{e^2}\;.
\label{eq:tau_def}
\end{eqnarray}
It combines the topological angle $\theta$ related to instanton effect with the gauge coupling constant $e$ into such a single complex parameter. The holomorphy of $\tau$ implies that it only depends on the chiral superfield $A$ but not on its conjugate $\bar{A}$, i.e., $\tau=\tau(A)$. Indeed, $\tau$ can be regarded as a modular group parameter of $SL(2,\mathbb{Z})$, and its transformation properties are described by the $SL(2,\mathbb{Z})$ symmetry
\begin{eqnarray}
    \tau\to\tau'=\frac{a\tau+b}{c\tau+d}\;,\quad\mathcal{M}=\left(\begin{matrix}
        a & b\\
        c & d
    \end{matrix}\right)\in \text{SL}(2,\mathbb{Z})\;,
\end{eqnarray}
where $ad-bc=1$.
Under a duality transformation, the superfield components $(A_D,A)$ transform as a vector under $SL(2,\mathbb{Z})$
\begin{eqnarray}
    \left(\begin{matrix}
        A_D\\
        A
    \end{matrix}\right)\to\mathcal{M}\left(\begin{matrix}
        A_D\\
        A
    \end{matrix}\right)\;.
\end{eqnarray}
To keep the BPS mass formula $M_\text{BPS}(g,q)=|A^v_D(u)g+A^v(u)q|$ invariant, the charges $(g,q)$ of BPS states must transform accordingly as
\begin{eqnarray}
    \left(\begin{matrix}
        g'\\
        q'
    \end{matrix}\right)=\mathcal{M}^{-1~T}\left(\begin{matrix}
        g\\
        q
    \end{matrix}\right)\;.
\end{eqnarray}
This symmetry allows us to define the dual of $\tau$ as $\tau_D=\frac{\partial^2\mathcal{F}_D(A_D)}{\partial A_D^2}$. Under S-duality, the transformation is implemented by the $SL(2,\mathbb{Z})$ matrix with $a=d=0,b=1$ and $c=-1$. The dual coupling thus transforms as $\tau_D=-\frac{1}{\tau}$. In T-duality, the matrix takes $a=b=d=1$ and $c=0$. It leads to the transformation $\tau_D=\tau+1$, which shifts the topological angle by $2\pi$.

To determine the values of $\tau(u)$ and $\tau_D(u)$ at each point on the moduli space, we need the explicit form of the prepotentials $\mathcal{F}(A)$ and $\mathcal{F}_D(A_D)$. In SW theory, both the prepotential $\mathcal{F}$ and its dual prepotential $\mathcal{F}_D$ are completely determined by the SW curve in Eq.~\eqref{equ:epcurve}. Following the method in Ref.~\cite{Klemm:1995wp}, one can invert $A(u)$ ($A_D(u)$) into a series expansion in $u(A)$ ($u(A_D)$) and substitute it into the relation for $A_D(u)$ ($A(u)$) given in Eq.~\eqref{eq:defAAD}. Then, we perform the integral of $A$ or $A_D$.
In this way one can obtain the exact expressions of $\mathcal{F}$ and $\mathcal{F}_D$ as follows~\cite{Csaki:2024plt}
\begin{eqnarray}
    \mathcal{F}(A)&=&\frac{1}{2\pi i}\left\{-4A^2\left(\text{ln}\frac{2A}{\Lambda}-\frac{3}{2}\right)+A^2\sum_{k=1}^\infty d_k\left(\frac{\Lambda}{A}\right)^{4k}\right\}\;,
    \label{eq:FF}\\
    \mathcal{F}_D(A_D)&=&\frac{1}{4\pi i}\left\{A_D^2\left(\ln\frac{-iA_D}{32\Lambda}-\frac{3}{2}\right)-16iA_D+A_D^2\sum_{k=1}^\infty d_k^D\left(\frac{A_D}{\Lambda}\right)^k\right\}\;.
    \label{eq:FFD}
\end{eqnarray}
The perturbative parts of the prepotential $\mathcal{F}(A)$ and its dual $\mathcal{F}_D(A_D)$ arise from one-loop contributions of massive gauge multiplets and light monopoles, respectively. The series terms in the prepotentials correspond to non-perturbative corrections, with coefficients $d_k$ and $d_k^D$ referred to as ``instanton numbers''. Here we take the scenario of $\mathcal{N}=2$ $SU(2)$ supersymmetric theory with zero massless hypermultiplet and follow the procedure in Ref.~\cite{Klemm:1995wp} to calculate the instanton coefficients $d_k$ and $d_k^D$. We show the first 10 values of them in Appendix~\ref{app:ins}. There exist more calculation methods for the instanton coefficients of the effective prepotentials in Refs.~\cite{Ito:1995ga,Chan:1999gj,Tachikawa:2013kta}.
In $\mathcal{N}=2$ supersymmetric theory, instanton effect is the only allowed non-perturbative contribution. In the weakly coupled region with $|u|\gg \Lambda^2$, $\mathcal{F}(A)$ is dominated by its perturbative part, whereas the non-perturbative effects of the dual prepotential $\mathcal{F}_D(A_D)$ become significant. By contrast, in the strongly coupled region with $|u|\sim 2\Lambda^2$, the instanton terms in $\mathcal{F}(A)$ and the perturbative part of $\mathcal{F}_D(A_D)$ become the dominant contributions. It implies that in the strong coupling region, duality transformations map originally non-perturbative physics into a perturbatively accessible framework.

Differentiating Eqs.~\eqref{eq:FF} and \eqref{eq:FFD} yields the dependence of $\tau$ and $\tau_D$ on $A(u)$ and $A_D(u)$ as follows
\begin{eqnarray}
\tau(u)&=&\frac{\partial^2\mathcal{F}(A)}{\partial A^2}=\frac{1}{2\pi i}\left(-8\ln\frac{2A}{\Lambda}+\sum_{k=1}^\infty d_k(1-4k)(2-4k)\left(\frac{\Lambda}{A}\right)^{4k}\right)\;, \label{eq:tau}\\    \tau_D(u)&=&\frac{\partial^2\mathcal{F}_D(A_D)}{\partial A_D^2}=\frac{1}{4\pi i}\left(2\ln\frac{-iA_D}{32\Lambda}+\sum_{k=1}^\infty d_k^D(1+k)(2+k)\left(\frac{A_D}{\Lambda}\right)^k\right)\;.
\label{eq:tauD}
\end{eqnarray}
Based on the relation in Eq.~\eqref{eq:tau_def}, one can explicitly map out the dependence of $\theta(u)$ and $e(u)$ (or $\theta_D(u)$ and $e_D(u)$) on the modular parameter $u$. Specifically, we have
\begin{itemize}
    \item $|u|$ at large scale ($|u|\gg \Lambda^2$), $\displaystyle\tau(u)=\frac{4 i}{\pi}\ln{\frac{2A}{\Lambda}}\equiv\frac{\theta(u)}{2\pi}+\frac{4\pi i}{e(u)^2}$\;,
    \item $|u|$ near the scale ($|u|\sim 2\Lambda^2$), $\displaystyle \tau_D(u)=\frac{-i}{2\pi}\ln\frac{-iA_D}{32\Lambda}\equiv\frac{\theta_D(u)}{2\pi}+\frac{4\pi i}{e_D(u)^2}$\;.
\end{itemize}
To construct low-energy effective theory, such division of the moduli space aims to achieve analytical computation over non-perturbative effects within the framework of duality transformations. Concretely, in the weakly coupled region ($|u|\gg \Lambda^2$), only the photon remains massless in the low-energy spectrum. All other states such as charged vector bosons and magnetic monopoles acquire large masses and can be safely integrated out. This yields a perturbative $U(1)$ QED-like effective theory, referred to as the electric frame (E-frame). On the other hand, in the strongly coupled region near the singularity ($|u|\sim 2\Lambda^2$), the magnetic monopole, which is extremely heavy in the E-frame, now becomes massless or very light. One must then switch to the magnetic frame (M-frame), where the monopole is regarded as the new degree of freedom.
The transition between the two regions is precisely described by the $SL(2,\mathbb{Z})$ S-duality. This duality provides a ``window'' through which the originally non-perturbative, strongly coupled physics on the moduli space can be mapped onto perturbative descriptions. It thereby enables a complete characterization of the behaviour across the entire moduli space.

%%%%%%%%%%%%%%%%%%%%%%
\subsection{Seiberg-Witten axion with electric-magnetic duality}
%%%%%%%%%%%%%%%%%%%%%%

Next, we briefly review the SW axion model in Ref.~\cite{Csaki:2024plt}.
In SW theory, the multiplet also has a global $U(1)_\mathcal{R}$ symmetry. Its spontaneous breaking plays a role analogous to that of the PQ symmetry.
The analogous breaking of the $U(1)_\mathcal{R}$ symmetry gives rise to the particle called the $\mathcal{R}$-axion in the UV model.
However, in the low-energy effective theory, supersymmetry permits only one scalar degree of freedom, namely the scalar component of the $\mathcal{N}=1$ chiral superfield $A$ (or $A_D$).
This scalar field carries the $U(1)_\mathcal{R}$ charge 2.
Thus, the comlex phase induced by the symmetry breaking must be embedded into the field $A$. Moreover, $U(1)_\mathcal{R}$ is anomalous at the quantum level, with its chiral current $\partial_\mu j^\mu_\mathcal{R}\propto F_{\mu\nu}\tilde{F}^{\mu\nu}$. Consequently, similar to how the QCD axion acquires a mass from instanton background, this anomalous term also provides a non-perturbative potential for the IR axion and gives rise to an extremely light mass. The IR axion $a(x)$ now is defined as the complex phase of the scalar component $A(u)$
\begin{eqnarray}
A(x)=A^v(u)e^{i\frac{a(x)}{f(u)}}\;,
\end{eqnarray}
where $A^v(u)$ denotes the vev that depends on the moduli space corrdinate $u$ as discussed earlier, while $f(u)$ represents the axion decay constant. We will give its explicit definition later.

When the vev acquries an additional vacuum phase angle, the scalar field now contains two complex phases: one from the axion field $a(x)$ and the other from the vev phase angle itself, denoted as $\text{arg} A^v(u)$. We define the latter as a perturbative phase $\theta_p(u)=-8~\text{arg}A^v(u)$. After separating these two phases, we rewrite $A(x)$ as
\begin{eqnarray}
A(x)=|A^v(u)|e^{i\left(\text{arg}A^v(u)+\frac{a}{f}\right)}=|A^v(u)|e^{i\left(\frac{a}{f}-\frac{\theta_p}{8}\right)}=|A^v(u)|e^{i\alpha}\;,~~\alpha\equiv {a\over f}-{\theta_p\over 8}\;.
\end{eqnarray}
Therefore, Eq.~\eqref{eq:tau} can be rewritten to incorporate the axion as
\begin{eqnarray}
\tau&=&\frac{4i}{\pi}\ln\frac{2|A^v|}{\Lambda}-\frac{4\alpha}{\pi}-\frac{1}{2\pi}\sum_{k=1}^\infty(4k-1)(4k-2)d_k\left\vert\frac{\Lambda}{A}\right\vert^{4k}(\sin(4k\alpha)+i\cos(4k\alpha))\nonumber \\
&=&\frac{4\pi i}{e_p^2}-\frac{8\alpha+G(\alpha)}{2\pi}\;,
\label{eq:tau_reim}
\end{eqnarray}
where the perturbative coupling $e_p$ corresponds to the large $|u|$ regime, i.e., $e_p^2=\pi^2/\ln{(2|A^v|/\Lambda)}$.
From the expression of $\tau$ we have two conclusions. First, its real part $\text{Re}\tau$ consists of two terms. The term proportional to phase angle $\alpha$ contributes to the perturbative part of the axion-photon coupling $g_{a\gamma\gamma}$, while the other term from $G(\alpha)$ supplies its non-perturbative part. On the other hand, the imaginary part $\text{Im}\tau$ also contains two contributions. One corresponds to the perturbative term $1/e_p^2$ and the other comes from the imaginary part of $G(\alpha)$. Here $G(\alpha)$ encodes instanton effects, so that both $\text{Re}\tau$ and $\text{Im}\tau$ receive contributions from perturbative and non-perturbative parts.

Now let us examine how $\text{Re}\tau$ and $\text{Im}\tau$ enter the structure of the IR Lagrangian. In $\mathcal{N}=2$ supersymmetric SW theory, the IR Lagrangian is conventionally expressed as the sum of the kinetic terms for the chiral and vector superfields~\cite{Seiberg:1994aj,Seiberg:1994rs}
\begin{eqnarray}
\mathcal{L}_{\rm IR}=\frac{1}{8\pi i}\int d^4\theta\frac{\partial\mathcal{F}}{\partial A}\bar{A}+\frac{1}{8\pi i}\int d^2\theta\tau(A)W^\alpha W_\alpha+h.c.\;.
\end{eqnarray}
After expressing the Lagrangian in terms of the components of the vector boson $\partial_{[\mu,}V_{\nu]}\equiv F_{\mu\nu}$, we obtain
\begin{eqnarray}
\mathcal{L}_{\rm IR}=\frac{\text{Im}\tau}{4\pi}(\partial_\mu \bar{A})(\partial^\mu A)+\frac{\text{Im}\tau}{8\pi}F_{\mu\nu}F^{\mu\nu}+\frac{\text{Re}\tau}{8\pi}F_{\mu\nu}\tilde{F}^{\mu\nu}\;.
\label{eq:e_frame_L}
\end{eqnarray}
The first term corresponds to the scalar kinetic energy and is equivalent to %\TL{$Im\tau_D$?}
\begin{eqnarray}
\text{Im}\tau_D(\partial_\mu\bar{A}_D)(\partial^\mu A_D)/(4\pi)\;.
\end{eqnarray}
Accoding to the relations ${F_D}_{\mu\nu}=\text{Im}\tau F_{\mu\nu}+\text{Re}\tau\tilde{F}_{\mu\nu}{\tau}$ and $\tau_D=-\frac{1}{\tau}$ under the S-duality, one can see that the second and third terms in Eq.~\eqref{eq:e_frame_L}
are equivalent to
\begin{eqnarray}
\frac{\text{Im}\tau_D}{8\pi}{F_D}_{\mu\nu}{F_D}^{\mu\nu}+ \frac{\text{Re}\tau_D}{8\pi} {F_D}_{\mu\nu}\tilde{F}_D^{\mu\nu}\;.
\end{eqnarray}
Based on our previous analysis, $\tau$ and $\tau_D$ correspond to two frames that are S-dual to each other. We refer them to as the E-frame and the M-frame, respectively. Next, taking the electric description as an example, we substitute the real and imaginary parts of $\tau$ given in Eq.~\eqref{eq:tau_reim} into the Lagrangian of the E-frame in Eq.~\eqref{eq:e_frame_L}. The second and third terms can then be written as
\begin{eqnarray}
\mathcal{L}_{\rm IR}\supset -\left(-\frac{1}{2e_p^2}+\frac{\text{Im}G(\alpha)}{16\pi^2}\right)F_{\mu\nu}F^{\mu\nu}-\left(\frac{8\alpha+\text{Re}G(\alpha)}{16\pi^2}\right)F_{\mu\nu}\tilde{F}^{\mu\nu}\;.
\label{eq:axion-photon}
\end{eqnarray}
Obviously, the axion couples to the photon field $V_\mu$ in IR theory via two types of interactions: one is the CP-violating coupling of the form $aFF$, and the other is the conventional CP-conserving coupling $aF\tilde{F}$. However, we shall not investigate the electrodynamics of the SW axion directly from Eq.~\eqref{eq:axion-photon}. It is because the terms involving $G(\alpha)$ contain the axion field $a(x)$, and make the equations and their solutions exceedingly complicated. Instead, we expand $\tau$ around $a=0$
\begin{eqnarray}
\tau&=&\tau|_{a=0}+a\frac{\partial \tau}{\partial a}\bigg\vert_{a=0}+\cdots\;
\approx\tau_0+a\frac{\partial\tau}{\partial A}\frac{\partial A}{\partial a}\bigg\vert_{a=0}\nonumber\\
&=&\tau_0+\frac{ia}{f}|A^v|e^{i\text{arg} A^v}\frac{\partial\tau}{\partial A}\bigg\vert_{a=0}=\tau_0+2\pi \frac{g_{a\gamma\gamma}}{e^2}a\;.
\label{eq:tau_exp}
\end{eqnarray}
This expansion significantly simplifies the equations and allows the axion-photon coupling parameter $g_{a\gamma\gamma}$ to be written explicitly in the equations. The derivative of $\tau$ in Eq.~\eqref{eq:tau_exp} is
\begin{eqnarray}
\frac{\partial\tau}{\partial A}\bigg\vert_{a=0}&=&\frac{1}{2\pi i}\left(-8\frac{\Lambda}{2A}\frac{2}{\Lambda}+\sum_{k=1}^\infty 4k(4k-1)(4k-2)d_k\left(\frac{\Lambda}{A^v}\right)^{4k-1}\times\left(-\frac{\Lambda}{A^2}\right)\right)\;\nonumber\\
&=&-\frac{1}{2\pi iA}\left(8+\sum_{k=1}^\infty4k(4k-1)(4k-2)d_k\left(\frac{\Lambda}{A^v}\right)^{4k}\right)\;.
%={c_{a\gamma\gamma}\over 2\pi i A}\;.
\end{eqnarray}
After substituting this relation into Eq.~\eqref{eq:tau_exp} and defining a dimensionless constant $c_{a\gamma\gamma}$ as
\begin{eqnarray}
c_{a\gamma\gamma}=-8-\sum_{k=1}^\infty 4k(\tilde{b}_k-i\tilde{c}_k)\left(\frac{\Lambda}{A^v}\right)^{4k}\;,~(\tilde{b}_k-i\tilde{c}_k)=(4k-1)(4k-2)d_k\;,
\label{eq:cagamma}
\end{eqnarray}
we obtain the axion-photon coupling given by~\footnote{There are slightly different definitions of $c_{a\gamma\gamma}$ in the literature. For instance, Ref.~\cite{DiLuzio:2020wdo} uses $g_{a\gamma\gamma}=\frac{e^2}{8\pi^2 f_a}c_{a\gamma\gamma}$.
In this work, we follow the definition given in Ref.~\cite{Csaki:2024plt}.}
\begin{eqnarray}
g_{a\gamma\gamma}=\frac{e^2}{4\pi^2f}c_{a\gamma\gamma}=\frac{ie^3}{2\sqrt{2}\pi}\frac{A}{|A|}\left(\frac{\partial\tau}{\partial A^v}\right)\bigg\vert_{a=0}=\frac{ie^3}{2\sqrt{2}\pi}e^{i\text{arg}A^v}\left(\frac{\partial\tau}{\partial A^v}\right)\bigg\vert_{a=0}\;,
\end{eqnarray}
where $f\equiv\sqrt{2}|A^v|/e$. This definition ensures that the coefficient of the axion kinetic term in the original Lagrangian follows the conventional normalization $1/2$. In the next section we will show that the kinetic term for the scalar field is written as $\text{Im}\tau(\partial_\mu\bar{A})(\partial^\mu A)/(4\pi)$, where $\text{Im}\tau\supset \text{Im}[\theta/(2\pi)+i4\pi/e^2]=4\pi/e^2$. Hence, the kinetic term can be expanded as
\begin{eqnarray}
\frac{\text{Im}\tau}{4\pi}(\partial_\mu\bar{A})(\partial^\mu A)&\supset&\frac{1}{e^2}\vert\partial_\mu(A^v(u)e^{i\frac{a}f{}})\vert^2=\frac{1}{e^2}\vert A^v e^{i\frac{a}{f}}\times\frac{i}{f(u)}\partial_\mu a(x)\vert^2\nonumber\\
&=&\frac{1}{e^2}\frac{\vert A^v\vert^2}{f(u)^2}(\partial_\mu a)^2\equiv\frac{1}{2}(\partial_\mu a)^2\;.
\end{eqnarray}

%%%%%%%%%%%%%%%%%%%%%%%%%
\section{Solutions to Seiberg-Witten axion electrodynamics}
\label{sec:Solution}
%%%%%%%%%%%%%%%%%%%%%%%%%

In this section, we will use the expansion of $\tau$ derived in Sec.~\ref{sec:SW} to derive the EM equation of motion (EoM) with SW axion. However, we will not expand $\tau$ from the beginning. Instead, we retain the full expression of $\tau$ throughout so that the symmetry structure of Lagrangian can be preserved and the physical interpretation becomes transparent. One can see that the EoMs from expanding $\tau$ in Eq.~\eqref{eq:e_frame_L} are identical to those obtained from Ref.~\cite{Csaki:2024plt}. The SW Lagrangians in the two frames are given as
\begin{eqnarray}
    &&\mathcal{L}_{\rm IR}^E=\frac{\text{Im}\tau^{ij}}{4\pi}(\partial_\mu\bar{A})_i(\partial^\mu A)_j+\frac{\text{Im}\tau^{ij}}{8\pi}(F_{\mu\nu})_i(F^{\mu\nu})_j+\frac{\text{Re}\tau^{ij}}{8\pi}(F_{\mu\nu})_i(\tilde{F}^{\mu\nu})_j\;,
    \label{eq:LagrangianE}\\
    &&\mathcal{L}_{\rm IR}^M=\frac{\text{Im}\tau_D^{ij}}{4\pi}(\partial_\mu\bar{A}_D)^i(\partial^\mu A_D)^j+\frac{\text{Im}\tau_D^{ij}}{8\pi}({F_D}_{\mu\nu})_i(F_D^{\mu\nu})_j+\frac{\text{Re}\tau_D^{ij}}{8\pi}({F_D}_{\mu\nu})_i(\tilde{F}_D^{\mu\nu})_j\;,
    \label{eq:LagrangianM}
\end{eqnarray}
where ${F}_{\mu\nu}^i=\text{Im}\tau_D^{ij}{({F_D}_{\mu\nu})}_j+\text{Re}\tau_D^{ij}{(\tilde{F}_{D\mu\nu})}_j$ and ${F_D}_{\mu\nu}^i=\text{Im}\tau^{ij}{(F_{\mu\nu})}_j+\text{Re}\tau^{ij}{(\tilde{F}_{\mu\nu})}_j$. The two Lagrangians above are physically equivalent, only written in a form that makes the electric-magnetic duality manifest. Here $\tau$ and $\tau_D$ are a pair of group parameters satisfying the S-duality transformation $\tau_D=-1/\tau$.
They respectively lead to the following EoMs
\begin{eqnarray}
    &&\text{E-frame:}~(\text{Re}\tau)\partial_\mu\tilde{F}^{\mu\nu}+\partial_\mu(\text{Im}\tau)F^{\mu\nu}+\partial_\mu(\text{Re}\tau)\tilde{F}^{\mu\nu}+(\text{Im}\tau)\partial_\mu F^{\mu\nu}=0\;,\\
    &&\text{M-frame:}~(\text{Re}\tau_D)\partial_\mu\tilde{F}_D^{\mu\nu}+\partial_\mu(\text{Im}\tau_D)F_D^{\mu\nu}+\partial_\mu(\text{Re}\tau_D)\tilde{F}_D^{\mu\nu}+(\text{Im}\tau_D)\partial_\mu F_D^{\mu\nu}=0\;.
\end{eqnarray}
After plugging the Maxwell equations $\partial_\mu F_D^{\mu\nu}=0$ and $\partial_\mu F^{\mu\nu}=0$ into them, they reduce to~\footnote{They can also be obtained by substituting the relations $F_D=\text{Im}\tau F+\text{Re}\tau\tilde{F}$ and $F=\text{Im}\tau_DF_D+\text{Re}\tau_D\tilde{F}_D$ in Maxwell equations $\partial_\mu F_D^{\mu\nu}=0$ and $\partial_\mu F^{\mu\nu}=0$, respectively.}
\begin{eqnarray}
    &&\text{E-frame:}~(\text{Re}\tau)\partial_\mu\tilde{F}^{\mu\nu}+\partial_\mu(\text{Im}\tau)F^{\mu\nu}+\partial_\mu(\text{Re}\tau)\tilde{F}^{\mu\nu}=0\;,\\
    &&\text{M-frame:}~(\text{Re}\tau_D)\partial_\mu\tilde{F}_D^{\mu\nu}+\partial_\mu(\text{Im}\tau_D)F_D^{\mu\nu}+\partial_\mu(\text{Re}\tau_D)\tilde{F}_D^{\mu\nu}=0\;.
\end{eqnarray}
We find that in a theory with electric-magnetic duality, the Bianchi identity $\partial_\mu \tilde{F}^{\mu\nu}=0$ no longer holds in either the E-frame or the M-frame. The Maxwell's equations, $\partial_\mu F^{\mu\nu}$, however, still satisfy the conservation equation $\partial_\mu F^{\mu\nu}=0$ in the absence of external sources.
Since electric charges and currents can be well shielded in the laboratory and magnetic charges have not been observed, we assume throughout the following that no external electromganetic sources are present as shown in Table~\ref{tab:maxwell}.

\begin{table}[htbp]
\centering
\begin{tabular}{lll}
\hline
\textbf{Equations}       & \textbf{E-frame}                          & \textbf{M-frame}                          \\
\hline
Bianchi Identity        & $\partial_\mu\tilde{F}^{\mu\nu} \neq 0$           & $\partial_\mu\tilde{F}_D^{\mu\nu} \neq 0$      \\
                        % & $(d F^{(2)} \neq 0)$                     & $(dF^{(2)}_D \neq 0)$                    \\
Maxwell Equations ($j_e^\nu=j_m^\nu=0$)      & $\partial_\mu F^{\mu\nu} = 0$                  & $\partial_\mu F^{\mu\nu}_D = 0$                 \\
% (no external sources)   & $(d \ast F^{(2)} = 0)$                   & $(d \ast F^{(2)}_D = 0)$                  \\
\hline
\end{tabular}
\caption{Comparison of Maxwell equations in E-frame and M-frame.
}
\label{tab:maxwell}
\end{table}

%%%%%%%%%%%%%%%%%%%%%%
\subsection{Electric frame}
%%%%%%%%%%%%%%%%%%%%%%

\begin{figure}[htbp!]
\centering
\includegraphics[width=0.5\linewidth]{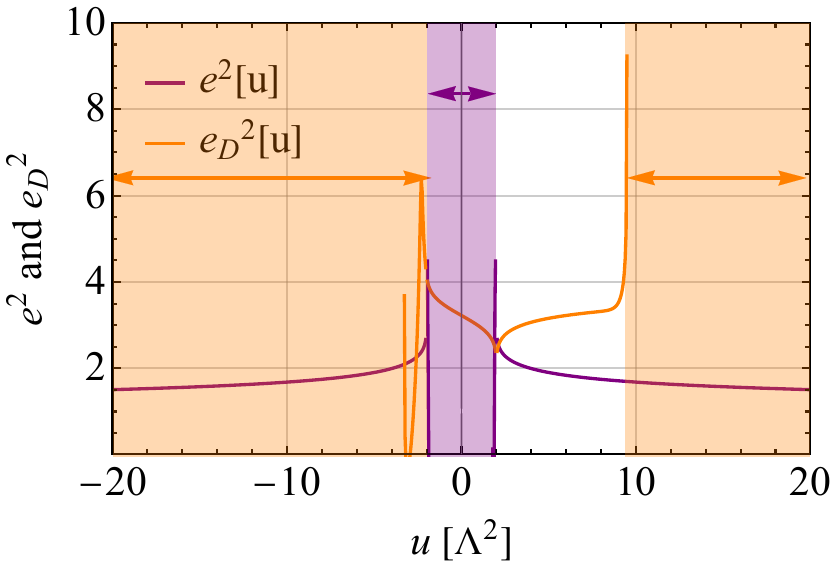}
\caption{The squared couplings $e^2(u)$ and $e_D^2(u)$ as a function of the coordinate $u$ in the electric frame and magnetic frame. 
}
\label{fig:e2eD2}
\end{figure}

We fisrt analyze the EoM in the electric frame. We decompose the field strength into a zeroth-order background field strength $F_0$ and an axion-induced field strength $F_a$, i.e., $F=F_0+F_a$. As the theory contains no external sources, the background field strength satisfies $\partial_\mu F_0^{\mu\nu}=0$ and $\partial_\mu \tilde{F}_0^{\mu\nu}=0$. The axion-induced field strength obeys $\partial_\mu F_a^{\mu\nu}=0$ and $\partial_\mu \tilde{F}_a^{\mu\nu}\neq0$. Consequently, the original equations can be further reduced to
\begin{eqnarray}
(\text{Re}\tau)\partial_\mu \tilde{F}_a^{\mu\nu}+\partial_\mu(\text{Im}\tau)F_0^{\mu\nu}+\partial_\mu(\text{Re}\tau)\partial_\mu \tilde{F}_0^{\mu\nu}=0\;.
\end{eqnarray}
Next, using the identity $\partial_\mu(\text{Re}\tau\tilde{F}_a^{\mu\nu})=\partial_\mu(\text{Re}\tau)\tilde{F}_a^{\mu\nu}+(\text{Re}\tau)\partial_\mu\tilde{F}_a^{\mu\nu}$, we absorb the coefficient $\text{Re}\tau$ in front of $\partial_\mu\tilde{F}_a^{\mu\nu}$ into the differential bracket. After introducing the redefined field $\mathbb{F}_a^{\mu\nu}\equiv (\text{Re}\tau)F_a^{\mu\nu}$, we obtain the field equation for $\mathbb{F}_a^{\mu\nu}$ as
\begin{eqnarray}
\partial_\mu\tilde{\mathbb{F}}_a^{\mu\nu}+\partial_\mu(\text{Im}\tau)F_0^{\mu\nu}+\partial_\mu(\text{Re}\tau)\tilde{F}_0^{\mu\nu}=0\;,
\label{eq:eom_simplified}
\end{eqnarray}
where we have dropped the term $\partial_\mu(\text{Re}\tau)\tilde{\mathbb{F}}_a^{\mu\nu}/{\text{Re}\tau}=\partial_\mu(\text{Re}\tau)\tilde{F}_a^{\mu\nu}$ since it is negligible compared to $\partial_\mu(\text{Re}\tau)\tilde{F}_0^{\mu\nu}$.
The equations can be rewritten in terms of the electromgantic fields $\mathbb{E}_a,\mathbb{B}_a,E_0$ and $B_0$ as follows
\begin{eqnarray}
&&\displaystyle\nabla\times\vec{\mathbb{E}}_a+\frac{\partial\vec{\mathbb{B}}_a}{\partial t}=\left(\vec{E}_0\times\nabla(\text{Re}\tau)-\vec{B}_0\frac{\partial(\text{Re}\tau)}{\partial t}\right)-\left(\vec{B_0}\times\nabla(\text{Im}\tau)+\vec{E}_0\frac{\partial(\text{Im}\tau)}{\partial t}\right)\;,
\label{eq:Maxwell1}\\
&&\nabla\cdot\vec{\mathbb{B}}_a=\displaystyle-\vec{B}_0\cdot\nabla(\text{Re}\tau)-\vec{E}_0\cdot\nabla(\text{Im}\tau)\;.
\label{eq:Maxwell2}
\end{eqnarray}
As both $\text{Re}\tau$ and $\text{Im}\tau$ contain the SW axion field, their specific values depend on the coordinate $u$ of the Coulomb branch. Here we restrict the parameter space to real values of $u$, yet not every point on the real axis yields smooth and positive electromagnetic couplings. Fig.~\ref{fig:e2eD2} shows the squared couplings $e^2(u)$ (purple line) and $e_D^2(u)$ (orange line) as a function of the coordinate $u$ in the two frames.
The purple and orange shadows respectively indicate the regions $|u|<2\Lambda^2$ for the E-frame coupling $e^2(u)$ and $u<-2\Lambda^2,u>9.39\Lambda^2$ for the M-frame coupling $e_D^2(u)$. They constitute singular regions which must be evaded when we work in E-frame or M-frame. For instance, coupling $e^2$ is not definitely positive in the interval $-2\Lambda^2<u<2\Lambda^2$.
In the range of $-2\Lambda^2<u<9.39\Lambda^2$, the M-frame charge $e_D^2(u)$ remains positive definite. Beyond this interval, $e_D^2(u)$ exhibits negative values or abrupt numerical jumps.

\begin{figure}
\centering
\includegraphics[width=0.5\linewidth]{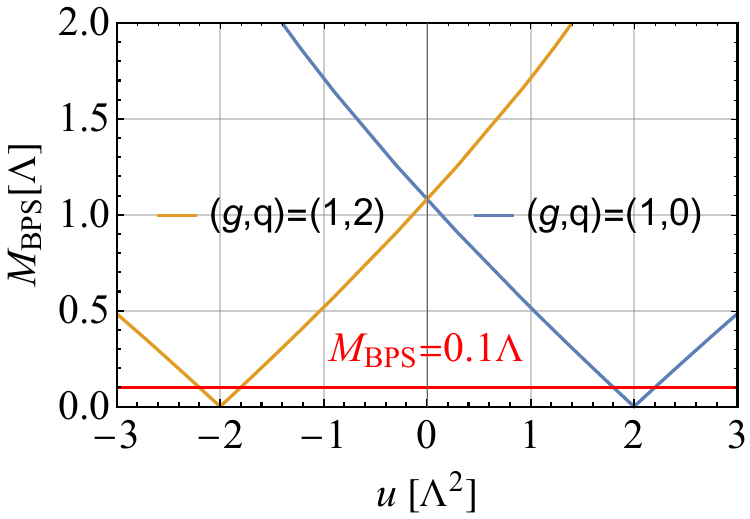}
\caption{BPS states mass spectrum on the Coulomb branch, illustrated for the monopole with $(g,q)=(1,0)$ and its first dyonic excitation with $(g,q)=(1,2)$. The masses of BPS states satisfy the central-charge formula $M_\text{BPS}=|A^v_D(u)g+A^v(u)q|$.
}
\label{fig:MBPS}
\end{figure}

One should note that the EoMs in Eqs.~\eqref{eq:Maxwell1} and \eqref{eq:Maxwell2} are derived from the IR Lagrangian given in Eq.~\eqref{eq:LagrangianE}. As discussed earlier, when $u$ approaches the singularities $u=\pm 2\Lambda^2$, two massless BPS states appear, namely monopole with $(g,q)=(1,0)$ and dyon with $(g,q)=(1,2)$. The SW-IR theory, however, only includes the vector superfield component $V_\mu$ as the physical photon and does not incorporate these light BPS states. To avoid these extra dynamical degrees of freedom, the BPS mass must not be too small so that we can integrate them out. By adopting the criterion $M_\text{BPS}\gtrsim 0.1\Lambda$, we obtain the excluded region $1.8\Lambda^2<|u|<2.2\Lambda^2$, as shown in Fig.~\ref{fig:MBPS}.

\begin{figure}
\centering
\includegraphics[width=0.49\linewidth]{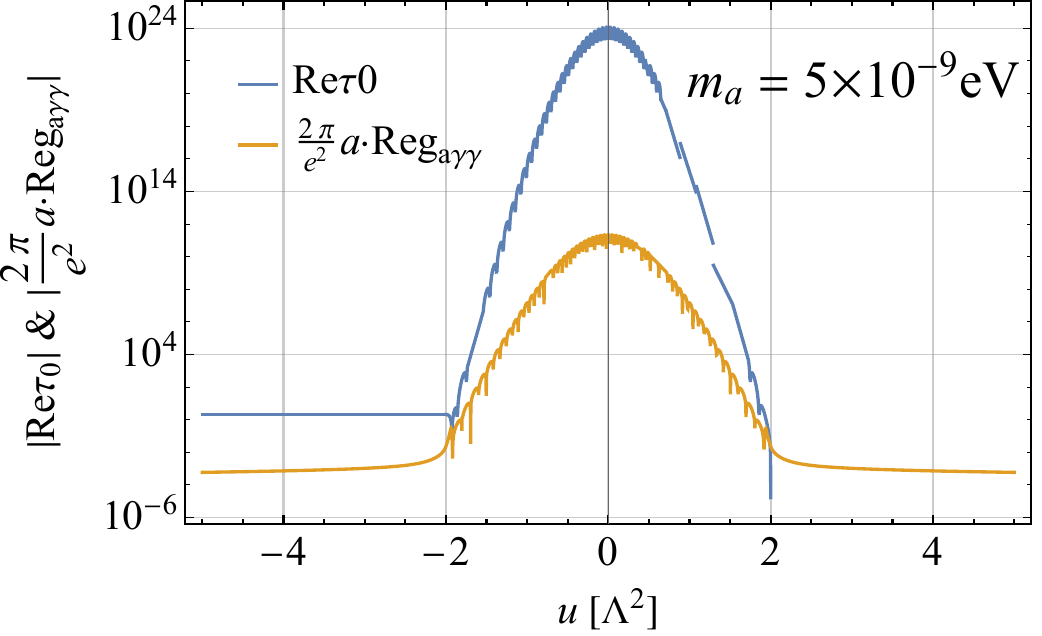}
\includegraphics[width=0.49\linewidth]{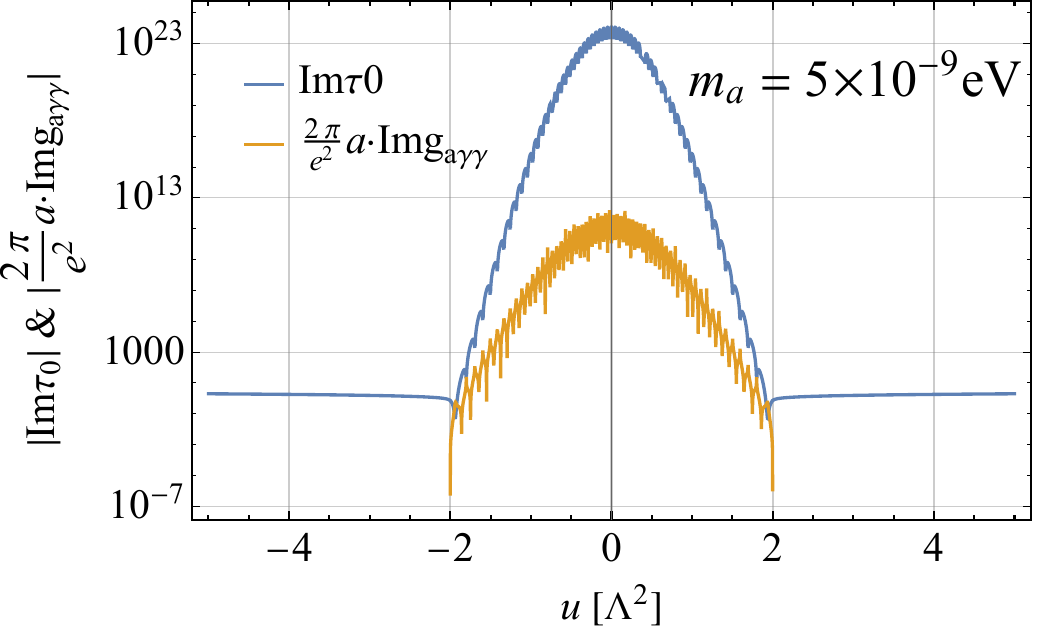}
\caption{Absolute values of the expansion terms for $\text{Re}\tau$ (left) and $\text{Im}\tau$ (right) on the real axis of $u$. The regions not shown here correspond to zero.
The axion mass is fixed to $m_a=5\times 10^{-9}~\text{eV}$.
}
\label{fig:Retau_Imtau}
\end{figure}

We could next take the curl on both sides of Eqs.~\eqref{eq:Maxwell1} and \eqref{eq:Maxwell2} to obtain the wave equations for the electromganetic fields $\mathbb{E}_a$ and $\mathbb{B}_a$. However, the present form of Eqs.~\eqref{eq:Maxwell1} and \eqref{eq:Maxwell2} is rather involved and not convenient for subsequent calculation. We therefore need to identify which terms can be safely neglected. As a first step in this simplification, we expand $\tau$ as follows
\begin{eqnarray}
\text{Re}\tau&\approx&\text{Re}[\tau_0+\frac{2\pi}{e^2}ag_{a\gamma\gamma}]=\text{Re}\tau_0+\frac{2\pi}{e^2}a\cdot \text{Re}g_{a\gamma\gamma}\;,
\label{eq:expansion1}\\
\text{Im}\tau&\approx&\text{Im}[\tau_0+\frac{2\pi}{e^2}ag_{a\gamma\gamma}]=\text{Im}\tau_0+\frac{2\pi}{e^2}a\cdot \text{Im}g_{a\gamma\gamma}\;.
\label{eq:expansion2}
\end{eqnarray}
This expansion approximates $\tau$ around $a=0$. The left and right panels of Fig.~\ref{fig:Retau_Imtau} display the absolute values of the two expnasion terms for $\text{Re}\tau$ and $\text{Im}\tau$, respectively. The left panel shows that on the real axis of $u$, the second term proportional to $\text{Re}g_{a\gamma\gamma}$ is always non-zero. The first term independent of axion field only takes non-zero values when $u<2\Lambda^2$. For $u>2\Lambda^2$, the leading term of the $\tau$ expansion vanishes, and the higher-order terms are extremely small. This makes the expansion useless in this region, and hence we exclude the domain $u>2\Lambda^2$. The right panel of Fig.~\ref{fig:Retau_Imtau} indicates that for the region of $u<-2\Lambda^2$, only the leading term independent of axion-photon coupling exists in the expansion of $\text{Im}\tau$.

\begin{figure}
\centering
\includegraphics[width=0.49\linewidth]{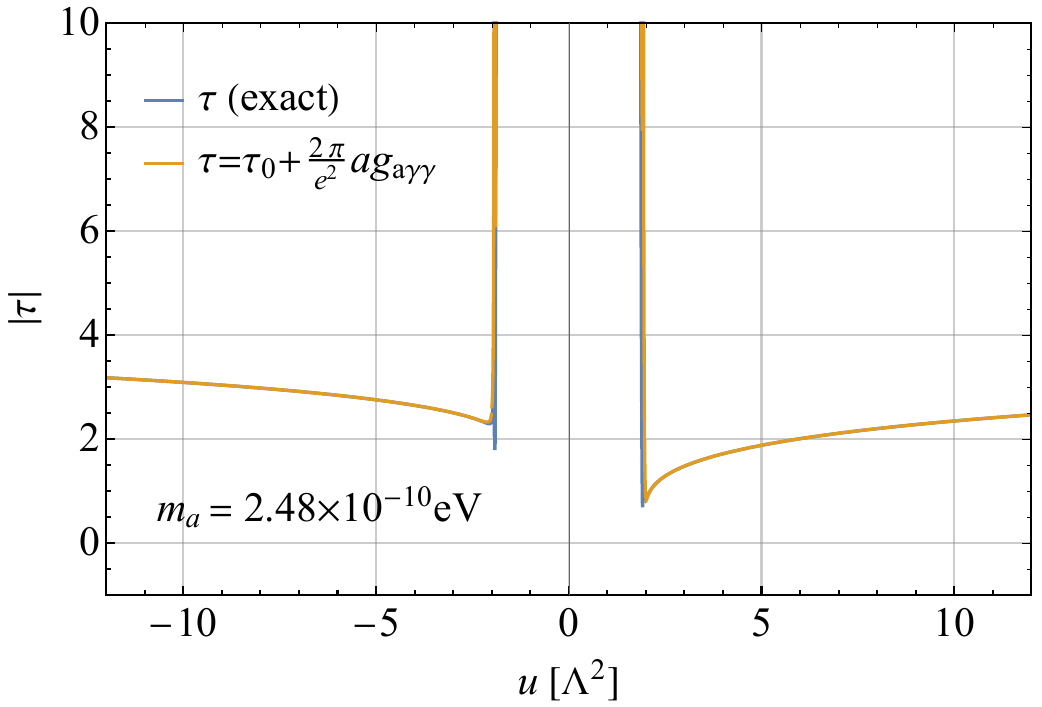}
\includegraphics[width=0.49\linewidth]{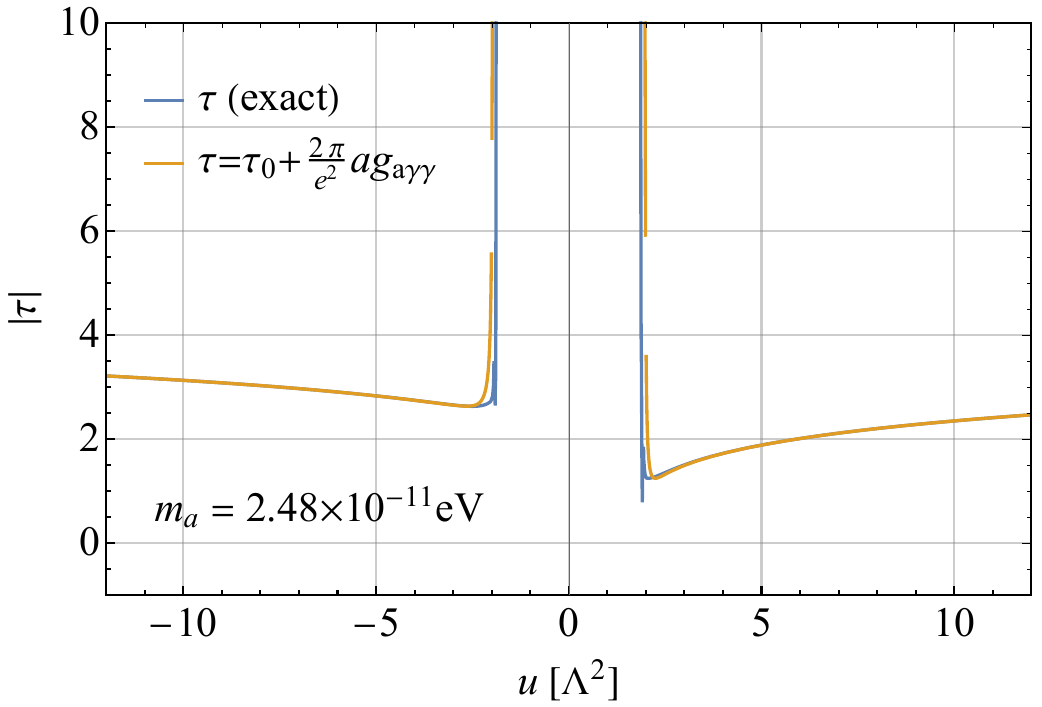}
\caption{Comparison between the exact $|\tau|$ (blue) and its approximate expansion (orange) for two different axion masses $m_a=2.48\times 10^{-10}~\text{eV}$ (left) and $2.48\times 10^{-11}~\text{eV}$ (right). }
\label{fig:tau}
\end{figure}

Another condition for the validity of the expansion in Eqs.~\eqref{eq:expansion1} and \eqref{eq:expansion2} is that $a\sim a_0=\sqrt{2\rho_{\rm DM}}/m_a$ with $\rho_{\rm DM}$ being the local DM density must not be too large. Thus, the axion mass $m_a$ cannot be too small. Fig.~\ref{fig:tau} illustrates the impact of two different mass choices on the validity of the expansion.
In the region of $-2\Lambda^2<u<2\Lambda^2$, the electric charge $e(u)$ has extremely small values and $|\tau|$ becomes very large. Thus, the expansion for $|u|<2\Lambda^2$ is not shown in Fig.~\ref{fig:tau}.
The left panel corresponds to $m_a=2.48\times 10^{-10}~\text{eV}$ (giving $a_0=0.01$). In this case, the exact form $\tau=\frac{4\pi i}{e_p^2}-\frac{8\alpha+G(\alpha)}{2\pi}$ (blue curve) coincides with the approximate form $\tau=\tau_0+\frac{2\pi}{e^2}ag_{a\gamma\gamma}$ (orange curve). The right panel clearly shows the discrepancy between the two, with $m_a=2.48\times 10^{-11}~\text{eV}$ (giving $a_0=0.1$).
To quantify the axion mass with which the approximate expansion of $\tau$ is reliable, we adopt the criterion that the relative difference between the exact and approximate forms at $|u|=2.2 \Lambda^2$ should be less than $0.1\%$. This yields $m_a>1.25\times 10^{-9}~\text{eV}$ for $u=-2.2\Lambda^2$ and $m_a>3.45\times 10^{-10}~\text{eV}$ for $u=2.2 \Lambda^2$.
In summary, the parameter space in the E-frame must satisfy
\begin{eqnarray}
    \boxed{{u<-2.2\Lambda^2,~m_a>1.25\times 10^{-9}~\text{eV}}}
\end{eqnarray}
This implies that within the above region, the terms $\nabla(\text{Im}\tau)$ and $\partial(\text{Im}\tau)/\partial t$ in the field equations can be directly dropped, while all terms involving $\text{Re}g_{a\gamma\gamma}$ are retained. Consequently, Eqs.~\eqref{eq:Maxwell1} and \eqref{eq:Maxwell2} are simplified to
\begin{eqnarray}
&&\nabla\times\vec{\mathbb{E}}_a+\frac{\partial\vec{\mathbb{B}}_a}{\partial t}=\vec{E}_0\times\nabla(\text{Re}\tau)-\vec{B}_0\frac{\partial(\text{Re}\tau)}{\partial t}\;,\\
&&\nabla\cdot \vec{\mathbb{B}}_a=0\;,
\end{eqnarray}
where we have utilized the relation of magnitude difference $\mathcal{O}(\nabla a(x,t))\ll\mathcal{O}(\frac{\partial}{\partial t}a(x,t))$ during the simplification. After taking the curl on both sides and using the other EoM $\partial_\mu F^{\mu\nu}_a=0\to\partial_\mu\mathbb{F}^{\mu\nu}_a=0$, we obtain the following wave equations
\begin{eqnarray}
    \nabla^2\vec{\mathbb{B}}_a-\frac{\partial^2\vec{\mathbb{B}}_a}{\partial t^2}&=&\vec{B}_0\frac{\partial^2\text{Re}\tau}{\partial t^2}-\vec{E}_0\times\nabla\frac{\partial\text{Re}\tau}{\partial t}\;,\label{eq:wave_eqs_E1}\\
    \nabla^2\vec{\mathbb{E}}_a-\frac{\partial^2\vec{\mathbb{E}}_a}{\partial t^2}&=&\frac{\partial\text{Re}\tau}{\partial t}\nabla\times\vec{B}_0-(\nabla\text{Re}\tau\cdot\nabla)\vec{E}_0\;.
    \label{eq:wave_eqs_E2}
\end{eqnarray}
Next we will solve the above equations~\cite{Sikivie_2021,Ouellet:2018nfr}. We consider two cases with only an external magnetic field $\vec{B}_0$ or only an external electric field $\vec{E}_0$.

%%%%%%%%%%%%
\subsubsection{Case I: $\vec{B}_0\neq0,\vec{E}_0=0$}
%%%%%%%%%%%%

In the case where only an external magnetic field $\vec{B}_0$ is present, the right-hand side of the wave equation contains only time derivatives of the axion field ($\text{Re}\tau\sim a(x,t)$)
\begin{eqnarray}
\nabla^2\vec{\mathbb{B}}_a-\frac{\partial^2\vec{\mathbb{B}}_a}{\partial t^2}&=&\vec{B}_0\frac{\partial^2\text{Re}\tau}{\partial t^2}\;,\label{eq:caseI_Eq1}\\
\nabla^2\vec{\mathbb{E}}_a-\frac{\partial^2\vec{\mathbb{E}}_a}{\partial t^2}&=&\frac{\partial\text{Re}\tau}{\partial t}\nabla\times\vec{B}_0\;.
\label{eq:caseI_Eq2}
\end{eqnarray}
Thus, these terms are independent of the spatial distribution of the axion $\nabla a(x)$ and are not suppressed by the DM relative velocity.
The static external magnetic field is aligned along the $z$ direction and is provided by a long solenoid of radius $R$. To solve Eqs.~\eqref{eq:caseI_Eq1} and \eqref{eq:caseI_Eq2}, we decompose them in cylindrical coordinates $(\rho,\phi,z)$. The external magnetic field can then be parameterized as $\vec{B}_0=\theta(R-\rho)B_0\hat{z}$. The axion field is given by $a(t,\vec{r})=a_0\cos(\omega_a t-\vec{k}_a\cdot \vec{r})$ with $\omega_a=m_a$ and $\vec{k}_a=m_a\vec{v}_a$.
The direction of axion field is parameterized in spherical coordinates with the angles $(\theta,\phi,\xi)$, following the setup established in Ref.~\cite{Li:2022oel}.

After imposing boundary conditions at $\rho'=0$, $\rho'=\infty$ and  $\rho'=\omega_a\rho=\omega_aR$, we obtain the solution of the induced field $\vec{\mathbb{B}}_a$ and $\vec{\mathbb{E}}_a$ as
\begin{eqnarray}
    \vec{\mathbb{B}}_a&=&\begin{cases}
        \Big[\frac{i\pi}{2}\frac{2\pi}{e^2}\text{Re}g_{a\gamma\gamma}a_0B_0\omega_aRH_1^\dagger(\omega_aR)J_0(\omega_a\rho)-\frac{2\pi}{e^2}\text{Re}g_{a\gamma\gamma}a_0B_0\Big]e^{i\omega_at}\hat{z},~\rho<R\;,\\
        \frac{i\pi}{2}\frac{2\pi}{e^2}\text{Re}g_{a\gamma\gamma}a_0B_0\omega_aRJ_1(\omega_aR)H_0^\dagger(\omega_a\rho)e^{i\omega_at}\hat{z},~\rho>R\;,
    \end{cases}\\
    &\approx&\begin{cases}
        -\frac{2\pi}{e^2}\text{Re}g_{a\gamma\gamma}a_0B_0\left[\frac{(\omega_aR)^2}{2}(\gamma'(\omega_aR)-\frac{1}{2})+\frac{\omega_a^2\rho^2}{4}\right]e^{i\omega_at}\hat{z},~\rho<R\;,\\
        -\frac{2\pi}{e^2}\text{Re}g_{a\gamma\gamma}a_0B_0\frac{(\omega_aR)^2}{2}\gamma'(\omega_a\rho)e^{i\omega_at}\hat{z},~~~~~~~~~~~~~~~~~~~~~~~\rho>R\;,\nonumber
    \end{cases}\\
    \vec{\mathbb{E}}_a&=&\begin{cases}
        \frac{\pi}{2}\frac{2\pi}{e^2}\text{Re}g_{a\gamma\gamma}a_0B_0\omega_aRH_1^\dagger(\omega_aR)J_1(\omega_a\rho)e^{i\omega_at}\hat{\phi},~\rho<R\;,\\
        \frac{\pi}{2}\frac{2\pi}{e^2}\text{Re}g_{a\gamma\gamma}a_0B_0\omega_aRJ_1(\omega_aR)H_1^\dagger(\omega_a\rho)e^{i\omega_at}\hat{\phi},~\rho>R\;.
    \end{cases}\\
    &\approx&\begin{cases}
        -i\frac{1}{2}\frac{2\pi}{e^2}\text{Re}g_{a\gamma\gamma}a_0B_0\omega_a\rho e^{i\omega_at}\hat{\phi},~~\rho<R\;,\\
        -i\frac{1}{2}\frac{2\pi}{e^2}\text{Re}g_{a\gamma\gamma}a_0B_0\omega_a\frac{R^2}{\rho}e^{i\omega_at}\hat{\phi},\rho>R\;.\nonumber
    \end{cases}
    \label{eq:caseI}
\end{eqnarray}
One can see that the above solutions can be written in terms of Bessel functions. Here $J_0,J_1$ denote the spherical Bessel functions of the first kind. $H_0^\dagger, H_1^\dagger$ are the spherical Hankel functions of the first kind and describe outgoing wave. $\gamma'(x)$ is defined as $\gamma'(x)=\ln(2/x)+\gamma-i\pi/2$ with the Euler-Mascheroni constant being $\gamma\approx 0.5772$.
The simplified result employs the long Compton wavelengths approximation, i.e., $\lambda_a=2\pi/m_a\gg R$ and thus $\rho'=\omega_aR\ll1$, which allows us to simplify the Bessel functions. The result shows that when only an external magnetic field is present, only the $\hat{z}$ ($\hat{\phi}$) component of induced magnetic field $\vec{\mathbb{B}}_a$ (electric field $\vec{\mathbb{E}}_a$) is dominant.

%%%%%%%%%%%%%%%%
\subsubsection{Case II: $\vec{E}_0\neq0,\vec{B}_0=0$}
%%%%%%%%%%%%%%%%

In the case where only an external electric field $\vec{E}_0$ is present, the right-hand side of the wave equations is suppressed by the axion velocity. The velocity of axion DM is given by  $\vec{v}_a=v_a(\sin\theta\cos(\xi-\phi),\sin\theta\sin(\xi-\phi),\cos\theta)$, with its typical magnitude being of order $\mathcal{O}(10^{-3})$. The wave equations become
\begin{eqnarray}
    \nabla^2\vec{\mathbb{B}}_a-\frac{\partial^2\vec{\mathbb{B}}_a}{\partial t^2}&=&-\vec{E}_0\times\nabla\frac{\partial\text{Re}\tau}{\partial t}\;,\\
    \nabla^2\vec{\mathbb{E}}_a-\frac{\partial^2\vec{\mathbb{E}}_a}{\partial t^2}&=&-(\nabla\text{Re}\tau\cdot\nabla)\vec{E}_0\;.
\end{eqnarray}
Analogous to Case I, we can also realize a static external electric field within a radius $R$ using a parallel plate capacitor, with $\vec{E}_0=E_0\theta(R-\rho)\hat{z}$. We then obtain the following solutions
\begin{eqnarray}
    \vec{\mathbb{B}}_a&=&\begin{cases}
        \Big[(-\frac{i}{12}k\pi(\omega_aR)^3H(\omega_aR)+k\pi M(\omega_aR))J_1(\omega\rho)+\frac{k\pi}{12}(\omega_a\rho)^3Y_1(\omega_a\rho)H(\omega_a\rho)\\
        ~~~-k\pi J_1(\omega_a\rho)M(\omega_a\rho)\Big]e^{i\omega_at}\hat{\phi},~\rho<R\;,\\
        -\frac{i}{12}k\pi(\omega_a R)^3H(\omega_a R)H_1^\dagger(\omega_a\rho)\hat{\phi},~\rho>R\;,
    \end{cases}\\
    &\approx&\begin{cases}
        k\Big[-\frac{i}{24}\pi\omega_a^4R^3\rho H(\omega_aR)+\frac{\pi}{2}\omega_a\rho M(\omega_a\rho)+\frac{1}{12}\omega_a^4\rho^4(\ln(\omega_a\rho)+\gamma-\frac{1}{2})H(\omega_a\rho)\\
        ~~~~-\frac{1}{6}\omega_a^2\rho^2H(\omega_a\rho)-\frac{\pi}{2}\omega_a\rho M(\omega_a\rho)\Big]e^{i\omega_a t}\hat{\phi}\;,~\rho<R\;,\\
        \frac{1}{12}k\Big[\omega_a^4R^3\rho(\gamma'(\omega_a\rho)-\frac{1}{2})H(\omega_aR)-2\omega_a^2\frac{R^3}{\rho}H(\omega_a R)\Big]e^{i\omega_a t}\hat{\phi}\;,~\rho>R\;,\nonumber
    \end{cases}\\
    \vec{\mathbb{E}}_a&=&\begin{cases}
        -\frac{\pi}{2}\frac{2\pi}{e^2}\text{Re}g_{a\gamma\gamma}a_0E_0v_a\omega_a R H_0^\dagger (\omega_aR)J_0(\omega_a\rho)\sin\theta\cos(2\phi-\xi)e^{i\omega_at}\hat{z},~\rho<R\;,\\
        -\frac{\pi}{2}\frac{2\pi}{e^2}\text{Re}g_{a\gamma\gamma}a_0E_0v_a\omega_aRJ_0(\omega_aR)H_0^\dagger(\omega_a\rho)\sin\theta\cos(2\phi-\xi)e^{i\omega_at}\hat{z},~\rho>R\;,
    \end{cases}
    \\&\approx&\begin{cases}
        -i\frac{2\pi}{e^2}\text{Re}g_{a\gamma\gamma}a_0E_0 v_a\omega_a R\Big[\gamma'(\omega_aR)\left(1-\frac{\omega_a^2\rho^2}{4}\right) \\
        +\frac{1}{4}(1-\gamma'(\omega_aR))(\omega_aR)^2\Big]\sin\theta\cos(2\phi-\xi)e^{i\omega_at}\hat{z},~\rho<R\;,\\
        -i\frac{2\pi}{e^2}\text{Re}g_{a\gamma\gamma}a_0E_0v_a\omega_aR\Big[\gamma'(\omega_a\rho)(1-\frac{\omega_a^2R^2}{4})\\
        +\frac{1}{4}(1-\gamma'(\omega_a\rho))(\omega_a\rho)^2\Big]\sin\theta\cos(2\phi-\xi)e^{i\omega_at}\hat{z},~\rho>R\;,\nonumber
    \end{cases}
\end{eqnarray}
where the coefficient $k$ is given by $k=-\frac{2\pi}{e^2}\text{Re}g_{a\gamma\gamma}a_0E_0v_a\sin\theta\cos(2\phi-\xi)$ for $\vec{\mathbb{B}}_{a,\phi}$. $Y_1(x)$ is the Bessel function of the second kind. $H(x)$ is the generalized hypergeometric function and $M(x)$ is the Meijer G function
\begin{eqnarray}
    H(x)&=&\text{HypergeometricPFQ}\left[\left\{\frac{3}{2}\right\},\left\{2,\frac{5}{2}\right\},-\frac{x^2}{4}\right]\;,\\
    M(x)&=& \text{MeijerG}\left[\{\{1\},\{0\}\},\left\{\left\{\frac{1}{2},\frac{3}{2}\right\},\{0,0\}\right\},\frac{x}{2},\frac{1}{2}\right]\;.
\end{eqnarray}
The induced magnetic field in Case II have both $\hat{\phi}$ and $\hat{\rho}$ components. Their solution forms are similar, and each is suppressed by the axion velocity $|\vec{v}_a|=v_a$, just as the induced electric field $\vec{\mathbb{E}}_{a,z}$ is. In the above expression we only show the $\vec{\mathbb{B}}_{a,\phi}$ component. The $\vec{\mathbb{B}}_{a,\rho}$ component can be simply obtained by replacing the coefficient $k$ with $-2\frac{2\pi}{e^2}\text{Re}g_{a\gamma\gamma}a_0E_0v_a\sin\theta\sin(2\phi-\xi)$.
Based on the results solved so far, the induced fields in the presence of $\vec{E}_0$ are all suppressed by $v_a$. This implies that, for the purpose of measuring $\text{Re}g_{a\gamma\gamma}$ within the E-frame, the most promising approach is employing an external magnetic field $\vec{B}_0$ to conduct the observation experiment.

\begin{figure}[htbp!]
\centering
\includegraphics[width=0.49\linewidth]{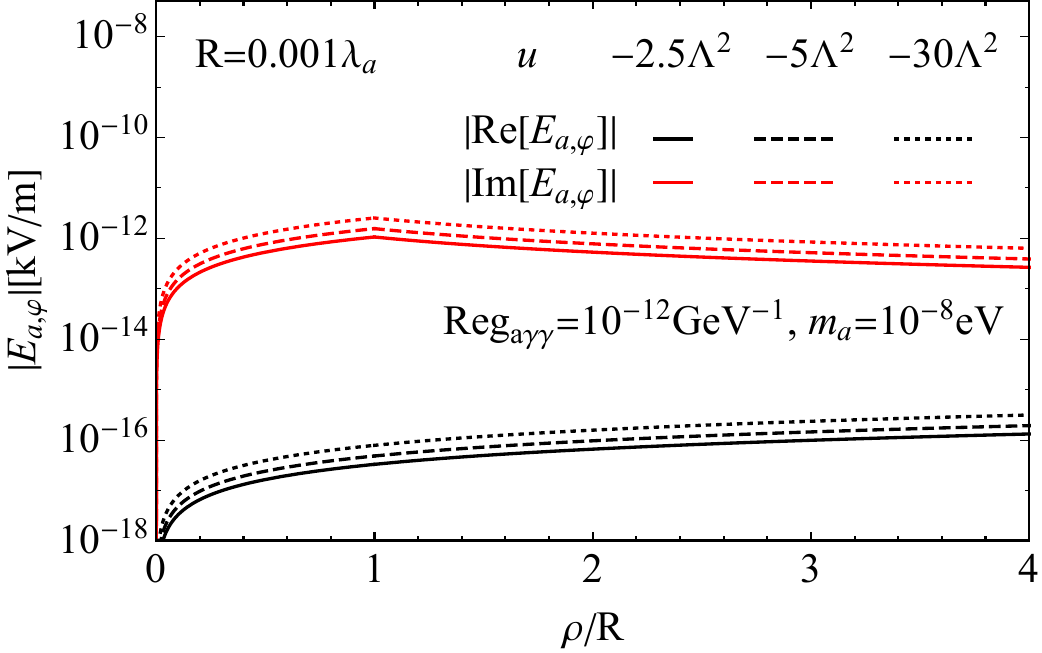}
\includegraphics[width=0.49\linewidth]{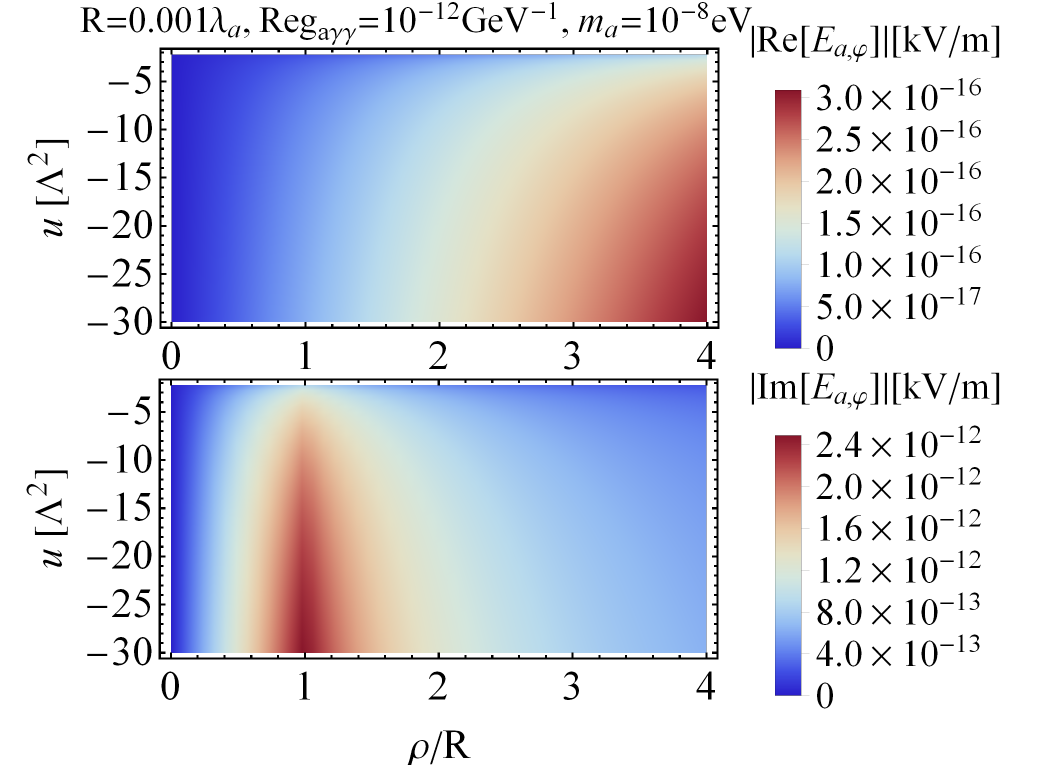}
\caption{Spatial profile of the induced electric field $|\vec{E}_{a,\phi}|$ along the direction $\hat{\phi}$ (in units of $\text{kV/m}$). The left panel shows results for three fixed values of $u$: $u=-2.5\Lambda^2$ (solid lines), $-5\Lambda^2$ (dashed lines) and $-30\Lambda^2$ (dotted lines). The right panel displays density plots over all $u$ satisfying $u<-2.2\Lambda^2$. Both panels adopt the same fixed parameters: $B_0=14~\text{T}$, $R=0.001\lambda_a$, $\text{Re}g_{a\gamma\gamma}=10^{-12}~\text{GeV}^{-1}$, and $m_a=10^{-8}~\text{eV}$.
}
\label{fig:induced_Ea}
\end{figure}

\begin{figure}[htbp!]
\centering
\includegraphics[width=0.49\linewidth]{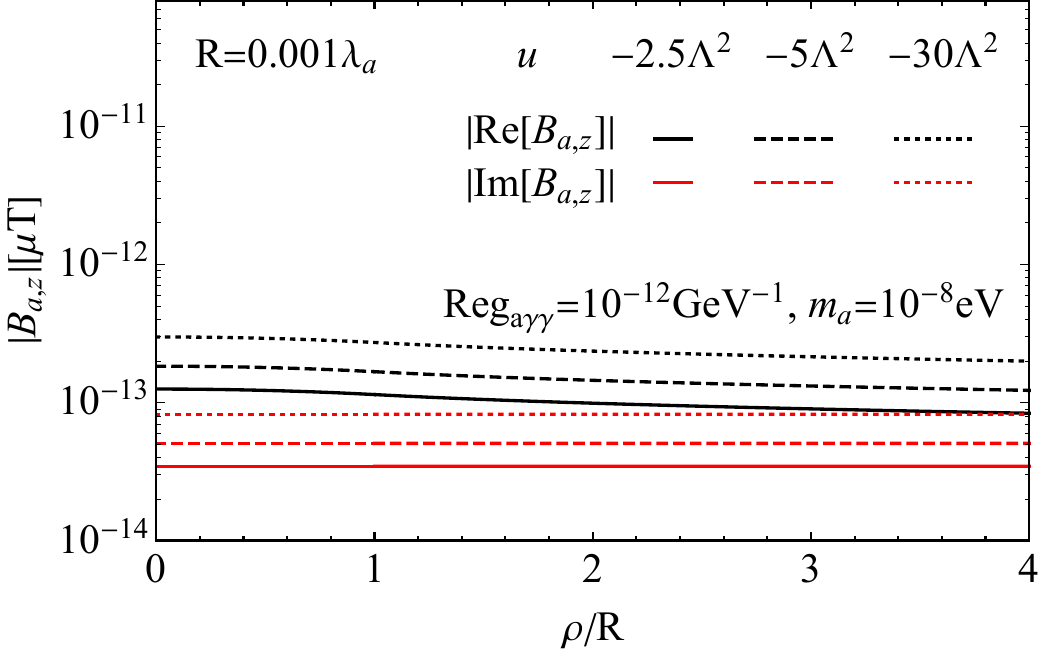}
\includegraphics[width=0.49\linewidth]{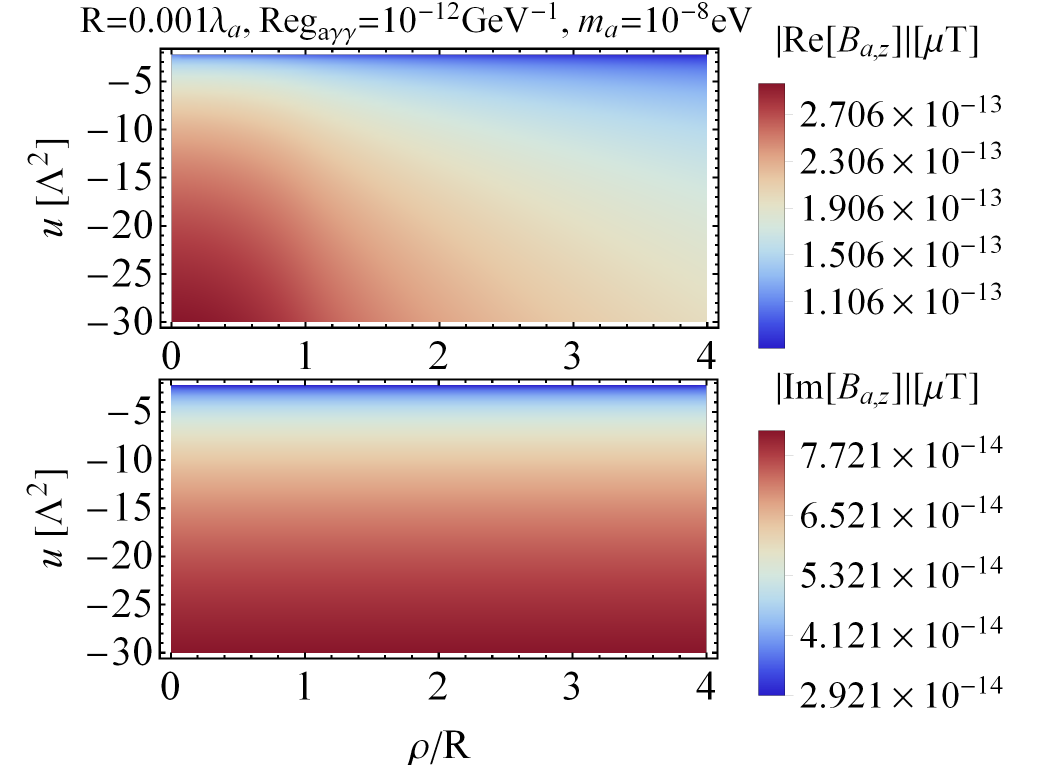}
\caption{Spatial profile of the induced magnetic field $|\vec{B}_{a,z}|$ along the direction $\hat{z}$ (in units of $\mu\text{T}$). The left panel shows results for three fixed values of $u$: $u=-2.5\Lambda^2$ (solid lines), $-5\Lambda^2$ (dashed lines) and $-30\Lambda^2$ (dotted lines). The right panel displays density plots over all $u$ satisfying $u<-2.2\Lambda^2$. Both panels adopt the same fixed parameters: $B_0=14~\text{T}$, $R=0.001\lambda_a$, $\text{Re}g_{a\gamma\gamma}=10^{-12}~\text{GeV}^{-1}$, and $m_a=10^{-8}~\text{eV}$.
}
\label{fig:induced_Ba}
\end{figure}

The field solutions $\vec{\mathbb{E}}_a$ and $\vec{\mathbb{B}}_a$ obtained above are not the physical electromagentic fields. To obtain the true physical field solutions $\vec{E}_a$ and $\vec{B}_a$, we must divide the above results by $\text{Re}\tau$
\begin{eqnarray}
    \mathbb{F}_a^{\mu\nu}=(\text{Re}\tau)F_a^{\mu\nu}\to \vec{E}_a=\vec{\mathbb{E}}_a/(\text{Re}\tau),~\vec{B}_a=\vec{\mathbb{B}}_a/(\text{Re}\tau)\;.
\end{eqnarray}
Since $\text{Re}\tau$ depends on the modular coordinate $u$, each value of $u$ on the Coulomb branch corresponds to a distinct field solution.
We specify the values of $u,m_a,B_0$ and $\text{Re}\tau$ for $u<-2.2\Lambda^2$. With a fixed relation between the spatial scale $R$ of $\vec{B}_0$ and the axion Compton wavelength $\lambda_a$, we can numerically compute the spatial profiles of the induced fields $|E_{a,\phi}|$ and $|B_{a,z}|$, as shown in Figs.~\ref{fig:induced_Ea} and \ref{fig:induced_Ba}, respectively. The left panels show the results for three fixed values of $u$: $u=-2.5\Lambda^2$ (solid lines), $-5\Lambda^2$ (dashed lines) and $-30\Lambda^2$ (dotted lines). The right panels display density plots over all $u$ satisfying $u<-2.2\Lambda^2$. Note that we have previously set a lower bound on $m_a$. In an actual measurement setup, an LC oscillator circuit has a minimum capacitance known as the stray capacitance. At resonance, where $\omega_a=1/\sqrt{LC}=m_a$, this minimal capacitance imposes an upper bound on $m_a$ that can be probed. Assuming typical values for the stray capacitance $C_{\text{stray}} = 50~\text{pF}$ and inductance $L = 10~\mu\text{H}$, the resonance condition yields an upper bound on the axion mass of $m_a = 2.95 \times 10^{-8}~\text{eV}$. Together with the previously established lower bound, the observable axion mass window is $1.25\times 10^{-9}~\text{eV}<m_a<2.95\times 10^{-8}~\text{eV}$.
We also fix the following parameters $B_0=14~\text{T}$, $R=0.001\lambda_a$, $\text{Re}g_{a\gamma\gamma}=10^{-12}~\text{GeV}^{-1}$, and $m_a=10^{-8}~\text{eV}$.

The numerical results in Figs.~\ref{fig:induced_Ea} and \ref{fig:induced_Ba} show that in the regime where $\lambda_a\gg R$ (illustrated by the case $R/\lambda_a=0.001$ in the plots), the size of induced EM field components becomes larger as the magnitude of $u$ increases. The induced electric field $\vec{E}_{a,\phi}$ can reach $10^{-12}~{\rm kV/m}$. To directly compare it with the induced magnetic field $\vec{B}_{a,z}$, we evaluate both fields at $\rho/R=1$. At this point with $u=-2.5\Lambda^2$, one can find $|\text{Im}[E_{a,\phi}]|=1.05\times 10^{-12}~\text{kV/m}$, while $|\text{Re}[B_{a,z}]|=3.41\times 10^{-14}~\text{kV/m}$ (converted from $\mu$T to \text{kV/m}). It turns out that the induced electric field $\vec{E}_{a,\phi}$ is two orders of magnitude larger than the induced magnetic field $\vec{B}_{a,z}$. Therefore, for the experiment employing an external magnetic field, a proper approach to probe the signal of axion field is to measure the induced electric field along the $\hat{\phi}$ direction, i.e., $\vec{E}_{a,\phi}$.
As seen in the right panel of Fig.~\ref{fig:induced_Ea}, $|\text{Im}[E_{a,\phi}]|$ always exhibits the maximal value at the point $\rho/R=1$ and $|\text{Re}[E_{a,\phi}]|$ increases along with raised $\rho$ for any $u<-2.2\Lambda^2$. However, the right panel of Fig.~\ref{fig:induced_Ba} indicates that $|\text{Im}[B_{a,z}]|$ does not depend on $\rho$ for any values of $u$ and the maximum of $|\text{Re}[B_{a,z}]|$ can be reached for small $\rho$ and large $|u|$.

%%%%%%%%%%%%%%%%%%%%%%
\subsection{Magnetic frame}
%%%%%%%%%%%%%%%%%%%%%%
We now turn to the M-frame. Since its EoMs are identical in form to those of the E-frame, we can directly write down the corresponding equation
\begin{eqnarray}
\partial_\mu\tilde{\mathbb{F}}_{D,a}^{\mu\nu}+\partial_\mu(\text{Im}\tau_D)F_{D,0}^{\mu\nu}+\partial_\mu(\text{Re}\tau_D)\tilde{F}_{D,0}^{\mu\nu}=0\;.
\label{eq:eom_simplified_m}
\end{eqnarray}
In the M-frame, the field strength is redefined as $\mathbb{F}_{D,a}^{\mu\nu}=(\text{Re}\tau_D)F_{D,a}^{\mu\nu}$. $F_D$ can likewise be split into a zeroth-order backround field and an axion-induced component $F_D=F_{D,0}+F_{D,a}$. The IR axion $a_D$ is embedded in the complex phase of the superfield scalar component $A_D(x)$
\begin{eqnarray}
A_D(x)=iA_D^v(u)e^{i\frac{a_D(x)}{f_D(u)}}=|A_D^v(u)|e^{i(\text{arg}A_D^v(u)+\frac{\pi}{2}+\frac{a_D}{f_D})}=|A_D^v(u)|e^{i\left(\frac{a_D}{f_D}-\frac{\theta_{D,p}}{8}\right)}=|A_D^v(u)|e^{i\alpha_D}\;,
\end{eqnarray}
where $A_D^v(u)$ is the vev as a function of the Coulomb branch coordinate $u$ in the M-frame. The perturbative phase is defined as $\theta_{D,p}=-\text{arg}A_D^v-\frac{\pi}{2}$, and the axion decay constant is $f_D=\sqrt{2}|A_D^v|/e_D$.
The axion field in the M-frame $a_D(x)$ is the phase of $A_D(x)$ and carries the same $\mathcal{R}$ charge as that in E-frame. Under an S-duality transformation, the scalar components obey the following transformation
\begin{eqnarray}
    (A_D,A)^T\to  \left(\begin{matrix}
        a &b\\
        c &d
    \end{matrix}\right)(A_D,A)^T=\left(
    \begin{matrix}
        0&1\\
        -1&0
    \end{matrix}
    \right)(A_D,A)^T=(A,-A_D)^T\;.
\end{eqnarray}
Then, the transformation $A\leftrightarrow A_D$ inevitably induces a corresponding exchange $a\leftrightarrow a_D$, which can be determined by
\begin{eqnarray}
    (A_D,A)\to(A,-A_D),~~A=A^v(x)e^{i\frac{a(x)}{f(u)}},~A_D=iA_D^v(x) e^{i\frac{a_D(x)}{f_D(u)}}\;,\nonumber\\
    \text{or}~\tau\to\tau_D=-1/\tau,~~\tau\approx\tau_0+\frac{2\pi}{e^2}ag_{a\gamma\gamma},\tau_D\approx\tau_{D,0}+\frac{2\pi}{e_D^2}a_D g_{a\gamma\gamma}^D\;.
\end{eqnarray}
Since both $\rho_\text{DM}$ and $m_a$ are invariant under the frame change, the amplitude $a_0$ is therefore a duality-invariant quantity. Hence, when evaluating the effects induced by the DM axion, we consider the same physcial amplitude in either frame, i.e., $a_{D,0}=a_0=\sqrt{2\rho_\text{DM}/m_a}$.

The modular parameter $\tau_D$ can also be expanded around $a_D=0$
\begin{eqnarray}
\tau_D&=&\tau_D\Big\vert_{a_D=0}+a_D\frac{\partial\tau_D}{\partial a_D}\Big\vert_{a_D=0}+...\approx \tau_{D,0}+a_D\frac{\partial\tau_D}{\partial A_D}\frac{\partial A_D}{\partial a_D}\Big\vert_{a_D=0}\;\\
&=&\tau_{D,0}-\frac{a_D}{f_D}|A_D^v|e^{i\text{arg}A_D^v}\frac{\partial \tau_D}{\partial A_D}\Big\vert_{a_D=0}=\tau_{D,0}+2\pi \frac{g_{a\gamma\gamma}^D}{e_D^2}a_D\;,\nonumber
\end{eqnarray}
where the derivative of $\tau_D$ with respect to $A_D^v$ is given by
\begin{eqnarray}
    \frac{\partial\tau_D}{\partial A_D^v}\Big\vert_{a_D=0}&=&\frac{1}{4\pi i}\left(\frac{2}{A_D^v}+\sum_{k=1}^\infty d_k^D(k+1)(k+2)\frac{k}{A_D^v}\left(\frac{A_D^v}{\Lambda}\right)^k\right)\;\nonumber\\
    &=&\frac{1}{2\pi i A_D^v}\left(1+\sum_{k=1}^\infty \frac{d_k^D}{2}(k+1)(k+2)k\left(\frac{A_D}{\Lambda}\right)^k\right)\;.
\end{eqnarray}
Following the convention adopted in E-frame, we define the S-dual coupling parameter in the M-frame as $g_{a\gamma\gamma}^D=\frac{e_D^2}{4\pi^2f_D}c_{a\gamma\gamma}^D$. The dimensionless coefficient $c_{a\gamma\gamma}^D$ is given by
\begin{eqnarray}
    c_{a\gamma\gamma}^D\equiv 1+\sum_{k=1}^\infty\frac{k}{2}(\tilde{b}_k^D-i\tilde{c}_k^D)\left(\frac{A_D}{\Lambda}\right)^k\;,~~(\tilde{b}_k^D-i\tilde{c}_k^D)=(k+1)(k+2)d_k^D\;,
    \label{eq:cagammaD}
\end{eqnarray}
and we have
\begin{eqnarray}
g_{a\gamma\gamma}^D=\frac{e_D^2}{4\pi^2f_D}c_{a\gamma\gamma}^D=\frac{ie_D^3}{2\sqrt{2}\pi}\frac{A_D}{|A_D|}\left(\frac{\partial \tau_D}{\partial A_D^v}\right)\Big\vert_{a_D=0}=-\frac{e_D^3}{2\sqrt{2}\pi}e^{i\text{arg}A_D^v}\left(\frac{\partial \tau_D}{\partial A_D^v}\right)\Big\vert_{a_D=0}\;.
\end{eqnarray}
Then, $\tau_D$ can be expanded as $\tau_D\approx \tau_{D,0}+2\pi \frac{g_{a\gamma\gamma}^D}{e_D^2}a_D$. Accordingly, Eq.~\eqref{eq:eom_simplified_m} can be expressed in terms of the electromagnetic fields $\mathbb{E}_{D,a},~\mathbb{B}_{D,a},~E_{D,0}$ and $B_{D,0}$. Clearly, since Eq.~\eqref{eq:eom_simplified_m} is identical in form to Eq.~\eqref{eq:eom_simplified}, the electromagnetic field equations derived from it must also coincide with Eqs.~\eqref{eq:Maxwell1} and \eqref{eq:Maxwell2}. Moreover, the parameters related to the SW theory in M-frame, such as $\tau_D,g_{a\gamma\gamma}^D$ and $f_D$ retain the same definitions as their couterparts in E-frame. In fact, the coupling $g_{a\gamma\gamma}$ in the E-frame can be expressed in terms of that in the M-frame
\begin{eqnarray}
g_{a\gamma\gamma}&=& \frac{ie^3}{2\sqrt{2}\pi} e^{i\text{arg}A^v}\left(\frac{\partial\tau}{\partial A^v}\right)=\frac{ie^3}{2\sqrt{2}\pi} e^{i\text{arg}A^v}\frac{1}{\tau_D^2}\frac{\partial \tau_D}{\partial A_D^v}\frac{\partial A_D^v}{\partial A^v}\;\\
&=&-\frac{ie^3}{2\sqrt{2}\pi}  e^{i\text{arg}A^v}\frac{1}{\tau_D^3}\frac{\partial\tau_D}{\partial A_D^v}=\frac{ie_D^3}{2\sqrt{2}\pi}\left(-\frac{|\tau_D|}{\tau_D}\right)^3e^{i\text{arg}A^v}\frac{\partial \tau_D}{\partial A_D^v}\nonumber\\
&=&g_{a\gamma\gamma}^D\times -i\left(-\frac{|\tau_D|}{\tau_D}\right)^3e^{i\text{arg}(A^v-A_D^v)}=g_{a\gamma\gamma}^D\left(-\frac{|\tau_D|}{\tau_D}\right)^3e^{i[\text{arg}(A^v-A_D^v)-\frac{\pi}{2}]}\;,\nonumber
\label{eq:ggD_relation}
\end{eqnarray}
where we utilize the charge duality relation $e=e_D|\tau_D|$. This shows that $g_{a\gamma\gamma}$ in the two frames differ only by a phase factor. Thus, they are physically equivalent and their squared moduli are invariant under the S-dual transforamtion~\cite{Csaki:2024plt}
\begin{eqnarray}
|g_{a\gamma\gamma}|^2=|g_{a\gamma\gamma}^D|^2\;.
\label{eq:equivalent}
\end{eqnarray}
However, the above conclusion holds only at the theoretical level. If one aims to detect the induced field singals in experiment, the resulting signals will break this invariance. We now proceed to analyze the constraints on the coupling $g_{a\gamma\gamma}^D$ in the M-frame. We first apply the external fields within the setup described in the E-frame. Then, we are able to measure the induced fields, and use these measurements to infer and constrain the parameters of the dual frame. An indirect procedure is required via the duality relation. We show the flowchart below
\begin{eqnarray}
    &&\boxed{\text{external field: }\vec{E}_0,\vec{B}_0}^{E}\to\boxed{\text{external field: }\vec{E}_0^D,\vec{B}_0^D}^M\to\boxed{\text{induced field: }\vec{\mathbb{E}}_{D,a},\vec{\mathbb{B}}_{D,a}}^M\nonumber\\
    &&\to\boxed{\text{induced field: }\vec{E}_{D,a},\vec{B}_{D,a}}^M\to\boxed{\text{induced field: }\vec{E_a},\vec{B}_a}^E\to\boxed{\text{constraint: }g_{a\gamma\gamma}^D}^E\nonumber
\end{eqnarray}

\begin{figure}[htbp!]
\centering
\includegraphics[width=0.49\linewidth]{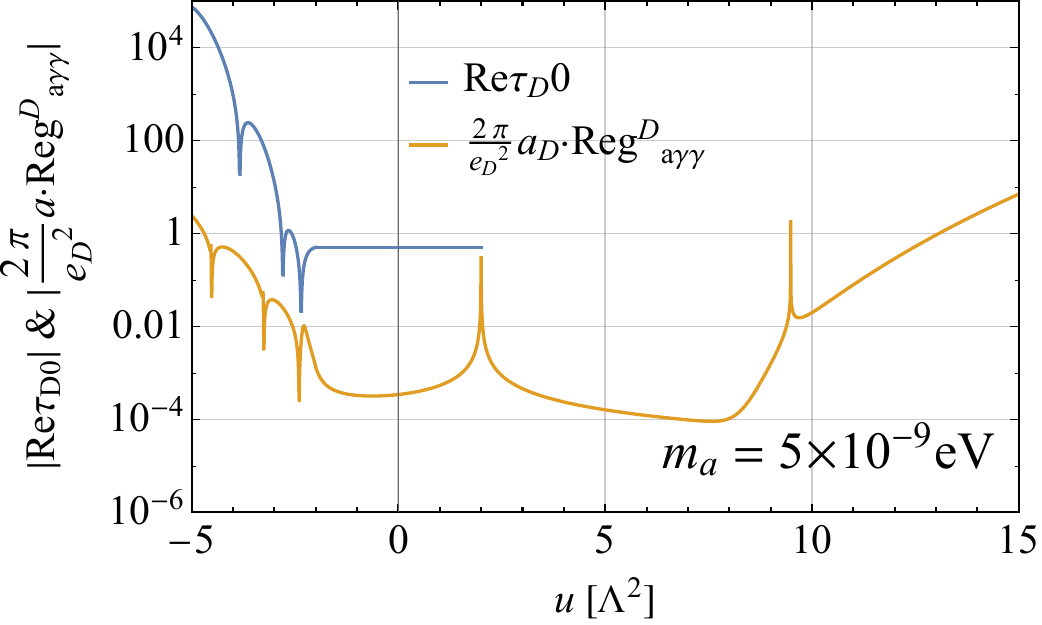}
\includegraphics[width=0.49\linewidth]{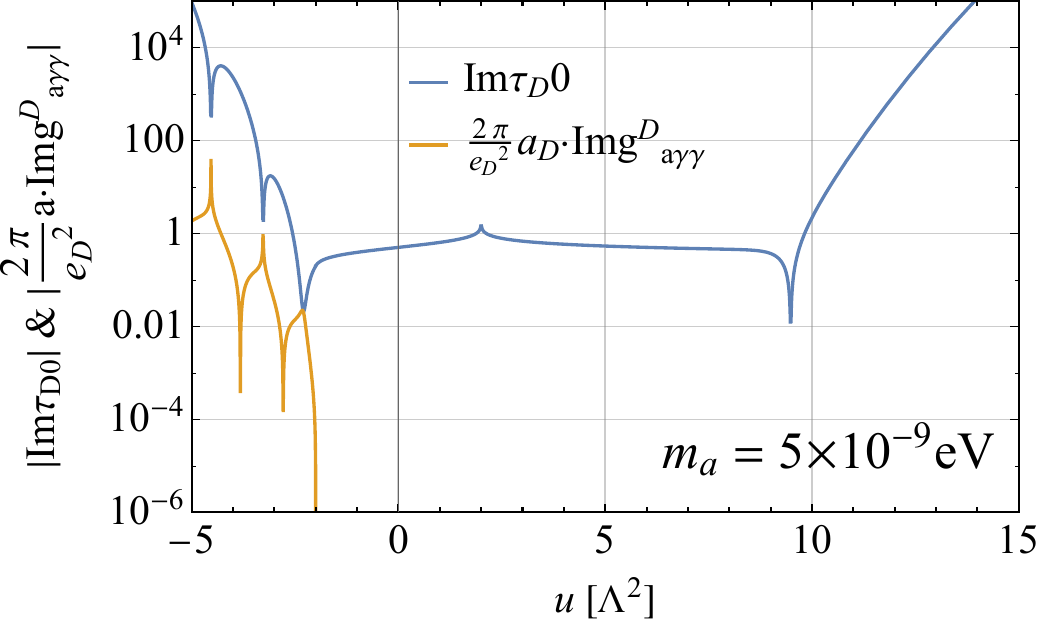}
\caption{
Absolute values of the expansion terms for $\text{Re}\tau_D$ (left) and $\text{Im}\tau_D$ (right) on the real axis of $u$. The regions not shown here correspond to zero.
The axion mass is fixed to $m_a=5\times 10^{-9}~\text{eV}$.
}
\label{fig:RetauD_ImtauD}
\end{figure}

To simplify the EoMs, we adopt the same strategy as in the E-frame by solving the equation within a resonable region consisting of the modulus $u$ and the axion mass $m_a$. According to the S-dual charge $e_D^2(u)$ shown in Fig.~\ref{fig:e2eD2}, this parameter yields smooth values only in the region $-2\Lambda^2<u<9.39145\Lambda^2$ on the real axis of $u$. The light BPS states impose the same exclusion of the region $1.8\Lambda^2<|u|<2.2\Lambda^2$. Fig.~\ref{fig:RetauD_ImtauD} shows the real (left) and imaginary (right) parts of the $\tau_D$. It turns out that, to ensure that the expansions of $\text{Re}\tau_D$ and $\text{Im}\tau_D$ remain reliable, the leading terms $\text{Re}\tau_{D,0}$ and $\text{Im}\tau_{D,0}$ must be non-zero. This requires an additional condition $u<2\Lambda^2$.

\begin{figure}[htbp!]
\centering
\includegraphics[width=0.49\linewidth]{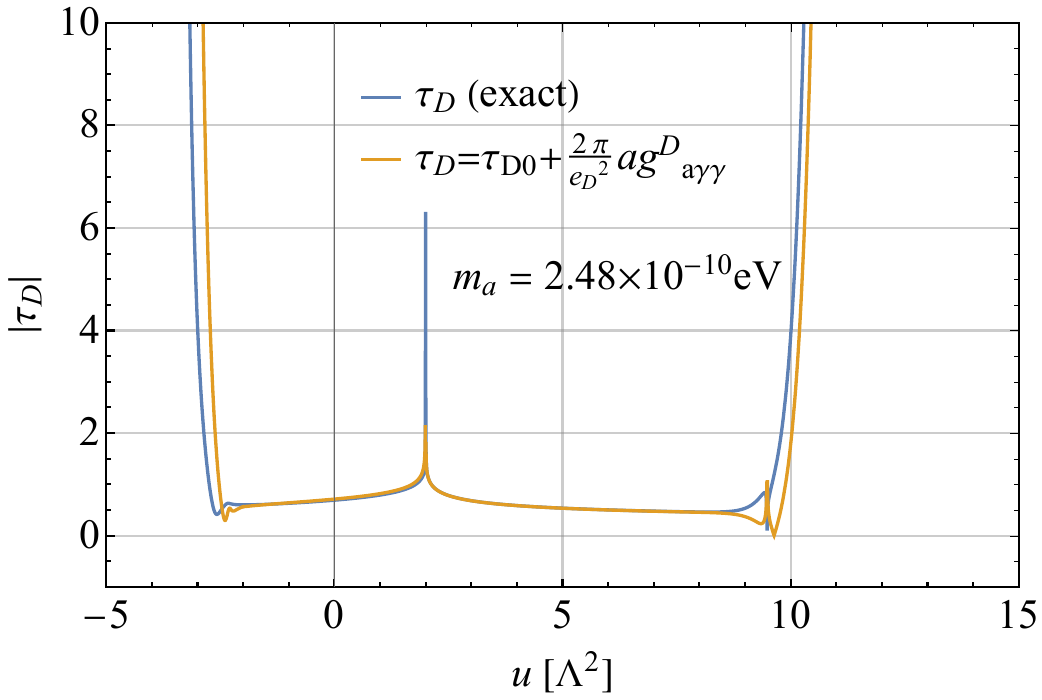}
\includegraphics[width=0.49\linewidth]{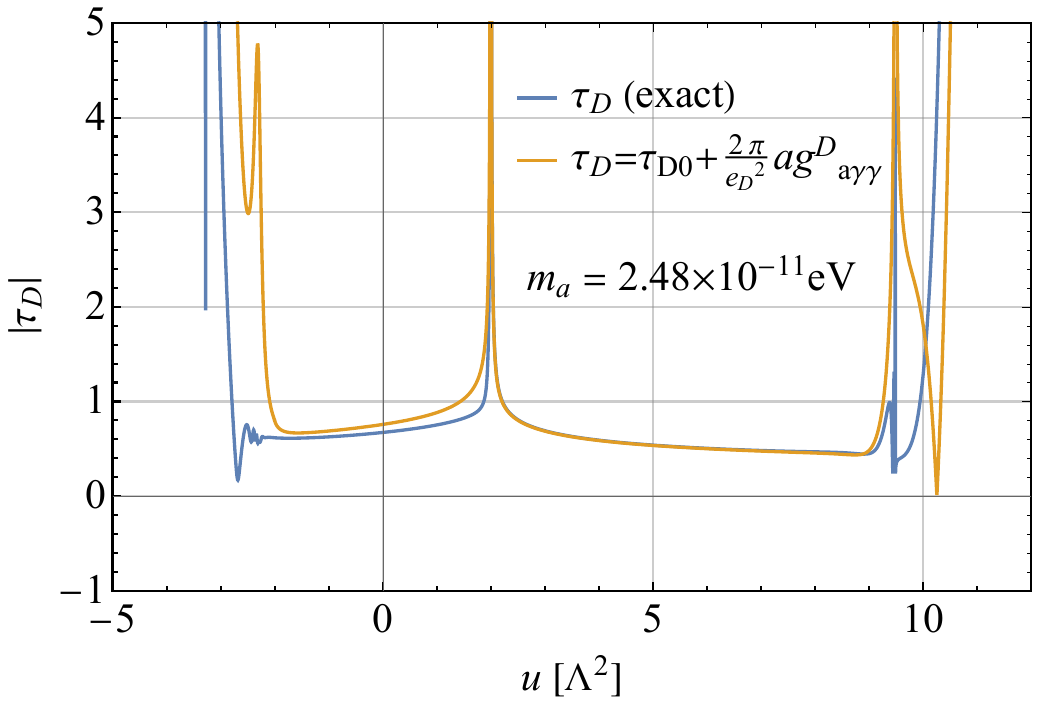}
\caption{
Comparison between the exact $|\tau_D|$ (blue) and its approximate expansion (orange) for two different axion masses $m_a=2.48\times 10^{-10}~\text{eV}$ (left) and $2.48\times 10^{-11}~\text{eV}$ (right).
}
\label{fig:tauD}
\end{figure}

Fig.~\ref{fig:tauD} illustrates the impact of different axion mass values on the accuracy of the approximate expansion for $\tau_D$. In M-frame, we consider that when $a_{D,0}<0.01$, i.e., $m_a>2.48\times 10^{-10}~\text{eV}$, the approximate form can safely agree with the exact one. In summary, the selected parameter space of $u$ and $m_a$ in the M-frame is
\begin{eqnarray}
\boxed{{-1.8\Lambda^2<u<1.8\Lambda^2\;\&~ u\neq0\;,~~m_a>2.48\times 10^{-10}~\text{eV}}}
\end{eqnarray}
Within this specified parameter region, we can get the wave equations that are totally identical in form to Eq.~\eqref{eq:wave_eqs_E1} and \eqref{eq:wave_eqs_E2}
\begin{eqnarray}
    \nabla^2\vec{\mathbb{B}}_{D,a}-\frac{\partial^2\vec{\mathbb{B}}_{D,a}}{\partial t^2}&=&\vec{B}_{D,0}\frac{\partial^2\text{Re}\tau_D}{\partial t^2}-\vec{E}_{D,0}\times\nabla\frac{\partial\text{Re}\tau_D}{\partial t}\;,\label{eq:wave_eqs_M1}\\
    \nabla^2\vec{\mathbb{E}}_{D,a}-\frac{\partial^2\vec{\mathbb{E}}_{D,a}}{\partial t^2}&=&\frac{\partial\text{Re}\tau_D}{\partial t}\nabla\times\vec{B}_{D,0}-(\nabla\text{Re}\tau_D\cdot\nabla)\vec{E}_{D,0}\;.
    \label{eq:wave_eqs_M2}
\end{eqnarray}
Before sloving the equations, it is important to note that all EM fields appearing in Eqs.~\eqref{eq:wave_eqs_M1} and \eqref{eq:wave_eqs_M2} are defined in the M-frame, whereas the fields we actually apply and detect in experiments are those of the E-frame. Using the duality relation $F_D=\text{Im}\tau F+\text{Re}\tau \tilde{F}$, the external fields in the E-frame can be transformed to them in the M-frame
\begin{eqnarray}
    B_{D,z,0}=\text{Im}\tau B_{z,0}+\text{Re}\tau E_{z,0}\;,~~
    E_{D,z,0}=\text{Im}\tau E_{z,0}-\text{Re}\tau B_{z,0}\;.
\end{eqnarray}
It is obvious to see that, regardless of whether only an external electric field or only an external magnetic field is present in the E-frame, both an electric field $E_{D,z,0}$ and a magnetic field $B_{D,z,0}$ are induced in the M-frame. Using the expansion $\text{Re}\tau_D=\text{Re}\tau_{D,0}+\frac{2\pi}{e_D^2}a_D \text{Re}g_{a\gamma\gamma}^D$, the wave equations Eqs.~\eqref{eq:wave_eqs_M1} and \eqref{eq:wave_eqs_M2} can be written as
\begin{eqnarray}
\nabla^2\vec{\mathbb{B}}_{D,a}-\frac{\partial^2\vec{\mathbb{B}}_{D,a}}{\partial t^2}&=&-\frac{2\pi}{e_D^2}\text{Re}g_{a\gamma\gamma}^D\vec{E}_{D,0}\times \vec{\nabla}\frac{\partial a_D}{\partial t}+\frac{2\pi}{e_D^2}\text{Re}g_{a\gamma\gamma}^D \frac{\partial ^2 a_D}{\partial t^2}\vec{B}_{D,0}\;,\\
\nabla^2\vec{\mathbb{E}}_{D,a}-\frac{\partial^2\vec{\mathbb{E}}_{D,a}}{\partial t^2}&=&-\frac{2\pi}{e_D^2}\text{Re}g_{a\gamma\gamma}^D(\vec{\nabla}a_D\cdot \nabla)\vec{E}_{D,0}+\frac{2\pi}{e_D^2}\text{Re}g_{a\gamma\gamma}^D\frac{\partial a_D}{\partial t}\vec{\nabla}\times \vec{B}_{D,0}\;.
\end{eqnarray}
We find that all solutions that depend on $E_{D,0}$ are suppressed.
Thus, in the M-frame the dominant solution depends only on $B_{D,0}$. As a result, its form becomes identical to that of Case I in E-frame, with all parameters simply replaced be their counterparts carrying the subscript $D$:
\begin{eqnarray}
\vec{\mathbb{B}}_{D,a}&=&\begin{cases}
\Big[\frac{i\pi}{2}\frac{2\pi}{e_D^2}\text{Re}g_{a\gamma\gamma}^Da_{D,0}B_{D,0}\omega_aRH_1^\dagger(\omega_aR)J_0(\omega_a\rho)-\frac{2\pi}{e_D^2}\text{Re}g_{a\gamma\gamma}^Da_{D,0}B_{D,0}\Big]e^{i\omega_at}\hat{z},~\rho<R\;,\nonumber\\
\frac{i\pi}{2}\frac{2\pi}{e_D^2}\text{Re}g_{a\gamma\gamma}^Da_{D,0}B_{D,0}\omega_aRJ_1(\omega_aR)H_0^\dagger(\omega_a\rho)e^{i\omega_at}\hat{z},~\rho>R\;,
\end{cases}\\
&\approx&\begin{cases}
-\frac{2\pi}{e_D^2}\text{Re}g_{a\gamma\gamma}^Da_{D,0}B_{D,0}\left[\frac{(\omega_aR)^2}{2}(\gamma'(\omega_aR)-\frac{1}{2})+\frac{\omega_a^2\rho^2}{4}\right]e^{i\omega_at}\hat{z},~\rho<R\;,\\
-\frac{2\pi}{e_D^2}\text{Re}g_{a\gamma\gamma}^Da_{D,0}B_{D,0}\frac{(\omega_aR)^2}{2}\gamma'(\omega_a\rho)e^{i\omega_at}\hat{z},~~~~~~~~~~~~~~~~~~~~~~~\rho>R\;,
\end{cases}\\
\vec{\mathbb{E}}_{D,a}&=&\begin{cases}
        \frac{\pi}{2}\frac{2\pi}{e_D^2}\text{Re}g_{a\gamma\gamma}^Da_{D,0}B_{D,0}\omega_aRH_1^\dagger(\omega_aR)J_1(\omega_a\rho)e^{i\omega_at}\hat{\phi},~\rho<R\;,\nonumber\\
        \frac{\pi}{2}\frac{2\pi}{e_D^2}\text{Re}g_{a\gamma\gamma}^Da_{D,0}B_{D,0}\omega_aRJ_1(\omega_aR)H_1^\dagger(\omega_a\rho)e^{i\omega_at}\hat{\phi},~\rho>R\;,
    \end{cases}\\
    &\approx&\begin{cases}
        -i\frac{1}{2}\frac{2\pi}{e_D^2}\text{Re}g_{a\gamma\gamma}^Da_{D,0}B_{D,0}\omega_a\rho e^{i\omega_at}\hat{\phi},~~\rho<R\;,\\
        -i\frac{1}{2}\frac{2\pi}{e_D^2}\text{Re}g_{a\gamma\gamma}^Da_{D,0}B_{D,0}\omega_a\frac{R^2}{\rho}e^{i\omega_at}\hat{\phi},\rho>R\;.
    \end{cases}
    \label{eq:magnetic_frame_solution}
\end{eqnarray}
The solutions obtained above are expressed in terms of the redefined field strength $\mathbb{F}_D$. To recover the physically observable EM field strength $F_D$, we need to divide the solution by $\text{Re}\tau_D$
\begin{eqnarray}
    \mathbb{F}_{D,a}^{\mu\nu}=(\text{Re}\tau_D)F^{\mu\nu}_a\to \vec{E}_{D,a}=\vec{\mathbb{E}}_{D,a}/(\text{Re}\tau_D),~~\vec{B}_{D,a}=\vec{\mathbb{B}}_{D,a}/(\text{Re}\tau_D)\;.
\end{eqnarray}
Since the forms of the induced field solutions in the two frames are identical, their relative numerical magnitudes should also obey the same pattern. The induced electric field $\vec{E}_{D,a}$ is larger than the induced magnetic field $\vec{B}_{D,a}$. For instance, at the point $\rho/R=1$ with $u=-1.5\Lambda^2$, we evaluate these two induced EM fields and find $|\text{Im}[E_{a,\phi}^D]|=1.07\times 10^{-12}~\text{kV/m}\gg |\text{Re}[B_{a,z}^D]|=3.50\times 10^{-14}~\text{kV/m}$ (converted from $\mu\text{T}$ to $\text{kV/m}$).
Thus, in this scenario, the detection strategy remains to measure the signal of the induced electric field $\vec{E}_{D,a}$ along the direction $\hat{\phi}$.

%%%%%%%%%%%%%%%%%%%%%%%%%%%%%%%%%%%%
\section{Prospective detection of Seiberg-Witten axion}
\label{sec:Sensitivity}
%%%%%%%%%%%%%%%%%%%%%%%%%%%%%%%%%%%%
The axion-induced electric field in $\hat{\phi}$ direction is a vortex field analogous to a Faradyay-induction field. It can be measured using a wire loop~\cite{Li:2022oel}. The detection setup has an LC circuit whose wire loop is placed inside a solenoid that provides the external magnetic field. When the wire loop picks up the induced field ${B}_{a,\phi}$ and the frequency of circuit satisfies $\omega_{LC}=\omega_a=m_a$, LC resonance occurs and amplifies the signal. The induced current in a loop of radius $R$ is given by
\begin{eqnarray}
    I_a=\frac{2\pi R E_{a,\phi}(R)}{R_s}\;,
\end{eqnarray}
where $a_0=\sqrt{2\rho_\text{DM}}/m_a$ with the local DM density being $\rho_{\text{DM}}=0.4~\text{GeV}~\text{cm}^{-3}$, and $R_s$ denotes the resistance $R_s=L\omega_a/Q_c$ with $Q_c$ being the quality factor of the LC circuit.

%%%%%%%%%%%%%%
\subsection{Electric frame}
%%%%%%%%%%%%%%

In the E-frame, when the axion-induced signal is detected, the LC resonant circuit yields a signal power
\begin{eqnarray}
    P_{\text{signal}}=\langle I_a^2R_s \rangle=\frac{4Q_c\pi^6 (\text{Re}g_{a\gamma\gamma})^2\rho_{\text{DM}}B_0^2R^4|H_1^\dagger(\omega_a R)J_1(\omega_a R)|^2}{L\omega_a e^4}\times\frac{1}{\vert\text{Re}\tau(g_{a\gamma\gamma})\vert^2}\;.
    \label{eq:power_E}
\end{eqnarray}
To measure this signal current, two typical experimental approaches are commonly employed. The first uses a highly sensitive SQUID magnetometer to directly detect the weak magnetic field generated by the signal current~\cite{Sikivie:2013laa}. The second feeds the signal current directly into an amplification chain for processing~\cite{Duan:2022nuy}. For the latter approach, the dominant noise includes thermal noise from the LC circuit and the noise introduced by the amplifier-detector chain. They can be described by an effective noise temperature $k_BT_N\approx 4k_B T_c+4k_B T_\text{Amp}$ with $T_ c$ and $T_\text{Amp}$ being the ambient temperature and the amplifier temperature, respectively. This leads to the noise power
\begin{eqnarray}
    P_\text{noise}=\kappa_B T_N\sqrt{\frac{\Delta f}{\Delta t}}\;,
\end{eqnarray}
where $\kappa_B$ is the Boltzmann constant, $\Delta f=f/Q_c$ is the detector bandwidth and $\Delta t$ is the observation time. To derive a constraint on the axion-photon coupling $\text{Re}g_{a\gamma\gamma}$, we require the signal-to-noise ratio (SNR) to satisfy
\begin{eqnarray}
    \text{SNR}=\frac{P_\text{signal}}{P_\text{noise}}>3\;.
\end{eqnarray}
The external magnetic field is set to $B_0=14~\text{T}$ and the circuit quality factor is $Q_c=10^4$~\cite{Sikivie:2013laa}. We take the observation time as $\Delta t=1~\text{week}$.

\begin{figure}
\centering
\includegraphics[width=0.49\linewidth]{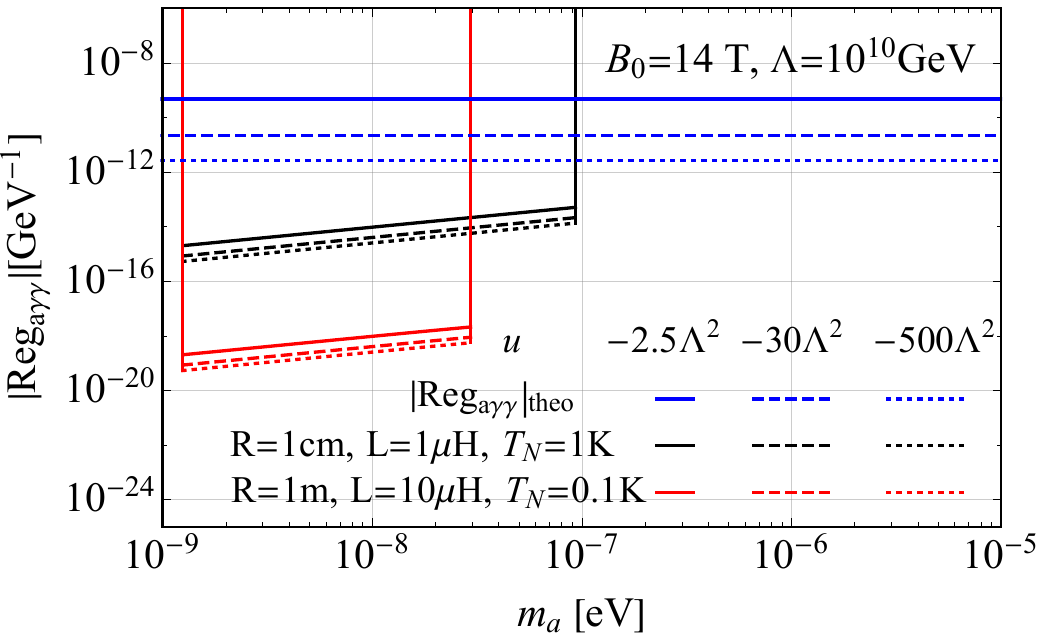}
\includegraphics[width=0.49\linewidth]{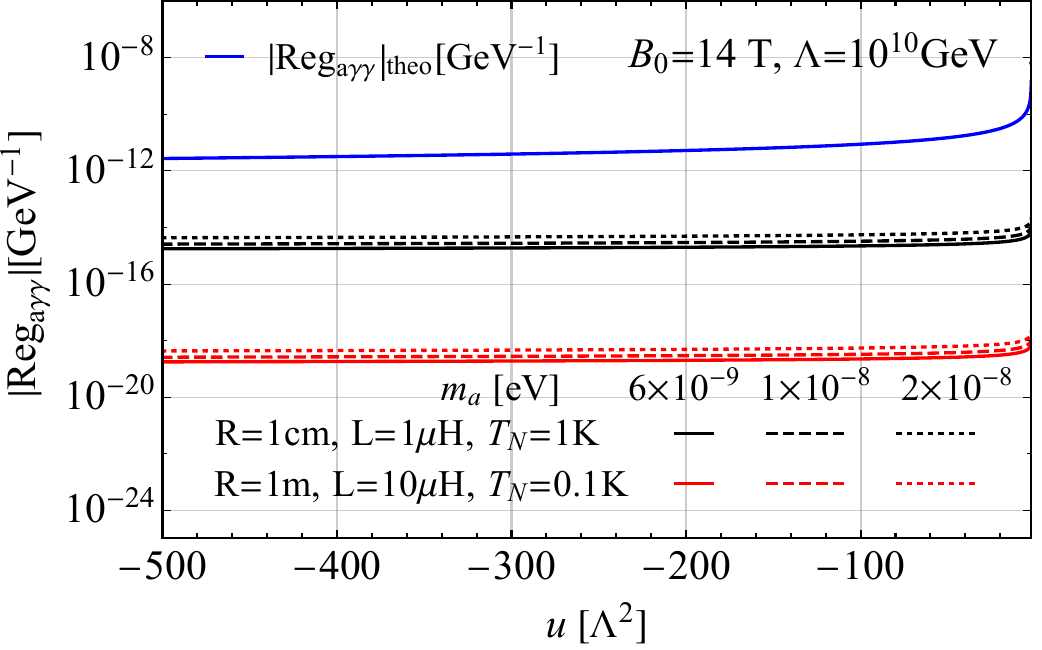}
\caption{The sensitivity limits on $|\text{Re}g_{a\gamma\gamma}|$ as a function of $m_a$ (left) or $u$ (right). Left: The value of $u$ is fixed as $u=-2.5\Lambda^2$ (solid lines), $-30\Lambda^2$ (dashed lines) or $-500\Lambda^2$ (dotted lines). Right: The value of axion mass $m_a$ is fixed as $m_a=6\times 10^{-9}$ eV (solid lines), $1\times 10^{-8}$ eV (dashed lines) or $2\times 10^{-8}$ eV (dotted lines). Two experimental setup benchmarks are considered: $R=1~\text{cm},L=1~\mu\text{H},T_N=1~\text{K}$ (black lines) and $R=1~\text{m},~L=10~\mu\text{H},~T_N=0.1~\text{K}$ (red lines). The predicted result of $|\text{Re}{g_{a\gamma\gamma}}|$ in theory is also displayed in blue lines. We adopt the SW scale $\Lambda=10^{10}~\text{GeV}$.
}
\label{fig:Reg}
\end{figure}

In Fig.~\ref{fig:Reg}, we present the sensitivity limits on $|\text{Re}g_{a\gamma\gamma}|$ as a function of $m_a$ (left) and $u$ (right). Two experimental setup benchmarks are considered: $R=1~\text{cm},L=1~\mu\text{H},T_N=1~\text{K}$ (black lines) and $R=1~\text{m},~L=10~\mu\text{H},~T_N=0.1~\text{K}$ (red lines). We adopt the SW scale $\Lambda=10^{10}~\text{GeV}$ for this estimate. The left panel shows that for a given axion mass, the sensitivity reach becomes stronger as the magnitude of $u$ increases, i.e., far from the singularity. For a fixed $u$ as shown in the right panel, the sensitivity becomes stronger for smaller axion masses. For instance, at $u=-500\Lambda^2$ and with the experimental configuration $R=1~\text{m},L=10~\mu\text{H},T_N=0.1~\text{K}$, the limit for $m_a=1.25\times 10^{-9}~\text{eV}$ is $|\text{Re}g_{a\gamma\gamma}|<5.26\times10^{-20}~\text{GeV}^{-1}$. Moreover, given our experimental setup benchmarks, the anticipated experimental sensitivity is expected to exceed the theoretical prediction shown in blue lines. The theoretical prediction of $\text{Re}g_{a\gamma\gamma}$ and $\text{Re}g_{a\gamma\gamma}^D$ are computed via the dimensionless coefficients Eqs.~\eqref{eq:cagamma} and \eqref{eq:cagammaD}, respectively, by including instanton contributions. %\TL{explain theoretical prediction a little bit}

%%%%%%%%%%%%%%%%
\subsection{Magnetic frame}
%%%%%%%%%%%%%%%%

\begin{figure}
\centering
\includegraphics[width=0.495\linewidth]{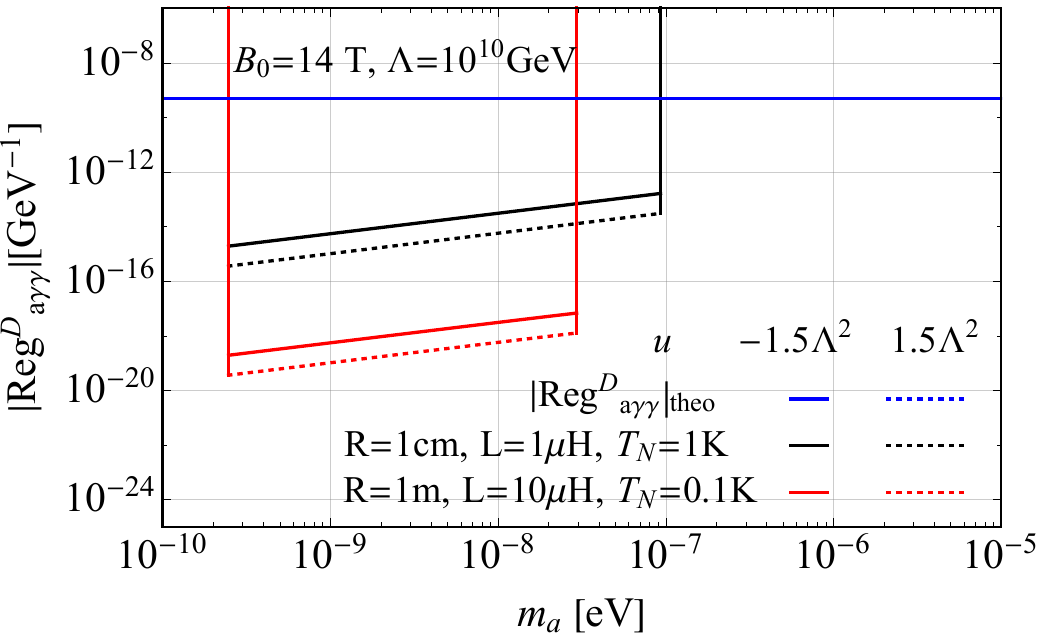}
\includegraphics[width=0.49\linewidth]{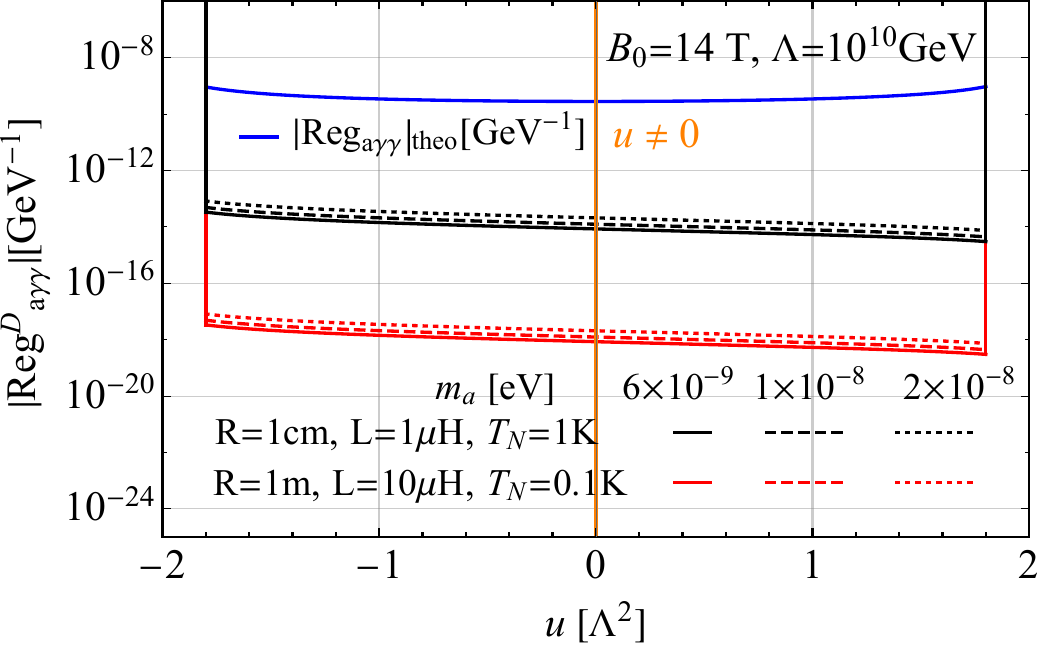}
\caption{
The sensitivity limits on $|\text{Re}g_{a\gamma\gamma}^D|$ as a function of $m_a$ (left) or $u$ (right). Left: The value of $u$ is fixed as $u=-1.5\Lambda^2$ (solid lines) or $1.5\Lambda^2$ (dotted lines). Right: The value of axion mass $m_a$ is fixed as $m_a=6\times 10^{-9}$ eV (solid lines), $1\times 10^{-8}$ eV (dashed lines) or $2\times 10^{-8}$ eV (dotted lines). Two experimental setup benchmarks are considered: $R=1~\text{cm},L=1~\mu\text{H},T_N=1~\text{K}$ (black lines) and $R=1~\text{m},~L=10~\mu\text{H},~T_N=0.1~\text{K}$ (red lines). The predicted result of $|\text{Re}{g_{a\gamma\gamma}^D}|$ in theory is also displayed in blue lines. We adopt the SW scale $\Lambda=10^{10}~\text{GeV}$.
}
\label{fig:RegD}
\end{figure}

By following the result of E-frame in Eq.~\eqref{eq:power_E}, we can write down the signal power related to the current induced by $\vec{E}_{D,a}$ in the M-frame as
\begin{eqnarray}
    P_\text{signal}^D=\langle I_{D,a}^2R_s\rangle=\frac{4Q_c\pi^6(\text{Re}g_{a\gamma\gamma}^D)^2\rho_\text{DM}^D B_{D,0}^2R^4|H_1^\dagger(\omega_aR)J_1(\omega_a R)|^2}{L\omega_a e_D^4}\times\frac{1}{|\text{Re}\tau_D(g_{a\gamma\gamma}^D)|^2}\;.
    \label{eq:power_M}
\end{eqnarray}
One can verify that the ratio between the signal powers of two frames is not equal to unity. This implies that the signal power is not a duality-invariant quantity. The reason lies in the fact that any actual measurement must pick a specific polarization direction. In this work, we detect only the $\hat{\phi}$-component of the induced electric field as mentioned before. Such a choice naturally breaks the original duality symmetry. We now translate the signal power $P_\text{signal}^D$ from the M-frame back into the E-frame, where the measurement is perfomed. Using the relation $F_{\mu\nu}=\text{Im}\tau_DF_{\mu\nu}^D+\text{Re}\tau_D\tilde{F}_{\mu\nu}^D$, we obtain
\begin{eqnarray}
\vec{E}_{a}=E_{a,\phi}=\text{Im}\tau_D\cdot E_{D,\phi}-\text{Re}\tau_D\cdot B_{D,\phi}\approx\text{Im}\tau_D\cdot E_{D,\phi}\;.
\end{eqnarray}
Thus, one can get the signal current power in the E-frame. This is the actual physical signal that can be observed experimentally
\begin{eqnarray}
P_\text{signal}&=&
\frac{4Q_c\pi^6(\text{Re}g_{a\gamma\gamma}^D)^2\rho_\text{DM}^D B_{0}^2R^4|H_1^\dagger(\omega_aR)J_1(\omega_a R)|^2}{L\omega_a e_D^4}\times\frac{|\text{Im}\tau|^2|\text{Im}\tau_D|^2}{|\text{Re}\tau_D|^2}\;\nonumber\\
&=&\frac{4Q_c\pi^6(\text{Re}g_{a\gamma\gamma}^D)^2\rho_\text{DM}^D B_{0}^2R^4|H_1^\dagger(\omega_aR)J_1(\omega_a R)|^2}{L\omega_a e_D^4}\times\frac{|\text{Im}\tau_D|^4}{|\tau_D|^4|\text{Re}\tau_D|^2}\;.
\end{eqnarray}
Similarly, imposing a requirement $\text{SNR}>3$ and setting the external magnetic field to $B_0=14~\text{T}$ yields prospective sensitivity limit on the coupling parameter $|\text{Re}g_{a\gamma\gamma}^D|$. Fig.~\ref{fig:RegD} shows the sensitivity limits on $|\text{Re}g_{a\gamma\gamma}^D|$ as a function of $m_a$ (left) and $u$ (right).
By measuring the signal converted back to the E-frame as $E_{a,\phi}=\text{Im}\tau_D\cdot E_{D,\phi}$, we obtain the limit $|\text{Re}g_{a\gamma\gamma}^D|<3.55\times 10^{-20}~\text{GeV}^{-1}$ at $m_a=2.48\times 10^{-10}~\text{eV}$ and $u=1.5~\Lambda^2$ for the experimental setup $R=1~\text{m},L=10~\mu\text{H},T_N=0.1~\text{K}$.

%%%%%%%%%%%%%%%%%%%%%%%%
\section{Conclusion}
\label{sec:Con}
%%%%%%%%%%%%%%%%%%%%%%%%

The $\mathcal{N}=2$ supersymmetric SW theory with its inherent $SL(2,\mathbb{Z})$ symmetry offers an ideal framework for us to study axion-monopole dynamics under electric-magnetic duality. The theory is determined by the geometry of the SW curve, and its prepotential clearly encodes non-perturbative instanton effect from magnetic monopoles. This allows us to examine non-quantized axion-photon couplings in instanton backgrounds.

In this work, starting from the IR Lagrangian of SW theory, we investigate the axion electrodynamics in supersymmetric SW theory. We first review the supersymmetric SW theory and SW axion model with electric-magnetic duality. The new EoMs respecting S-duality are then derived. We solve them in both electric frame and magnetic frame to obtain the axion-induced fields. We find that the dominant EM solution is the electric field in $\hat{\phi}$ direction $\vec{E}_{a, \phi}$ under an external magnetic field. The parameter space of moduli space coordinate $u$ is also analyzed in details. The E-frame is reliable for $u\lesssim -2\Lambda^2$, while the M-frame is valid for $|u|\lesssim 2\Lambda^2~(u\neq 0)$.
Based on these solutions, we propose experimental strategies to probe the axion-photon couplings in these two frames: $\text{Re}g_{a\gamma\gamma}$ and $\text{Re}g_{a\gamma\gamma}^D$. We show the prospective sensitivity of LC circuit to these axion-photon couplings and compare with theoretical predictions.

%###################################################################
%%%%%%%%%%%%%%%%%%%%%%%%
\acknowledgments
%%%%%%%%%%%%%%%%%%%%%%%%
T.~L. is supported by the National Natural Science Foundation of China (Grant No. 12375096, 12035008, 11975129).

%%%%%%%%%%%%%%%%%%%%%%%%%%%%%%%%%%%%%%%%%%%%%%
%%%%%%%%%%%%%%%%%%%%%%%%%%%%%%%%%%%%%%%%%%%%%%
\appendix
%%%%%%%%%%%%%%%%%%%%%%%%%%%%%%%%%%%%%%%%%%%%%%
%%%%%%%%%%%%%%%%%%%%%%%%%

%%%%%%%%%%%%%%%%%%%%%%%%%%%
\section{Instanton coefficients}
\label{app:ins}
%%%%%%%%%%%%%%%%%%%%%%%%%%%%%
To determine the exact form of the prepotentials $\mathcal{F}(A)$ and $\mathcal{F}_D(A_D)$, we follow the series inversion method in Ref.~\cite{Klemm:1995wp} to obtain the instanton coefficients. We used the first 30 instantons coefficients in our calculation. Here, we show the first ten coefficients of $d_k$ and its dual $d_k^D$ respectively
\begin{eqnarray}
    d_1&=&\frac{1}{2},~d_2=\frac{5}{64},~d_3=\frac{3}{64},~d_4=\frac{1469}{32768},~d_5=\frac{4471}{81920},~  d_6=\frac{40397}{524288}\;,\nonumber\\
    d_7&=&\frac{441325}{3670016},d_8=\frac{866589165}{4294967296},~d_9=\frac{383973403}{1073741824},~d_{10}=\frac{22730943821}{34359738368}\;,
\end{eqnarray}
and
\begin{eqnarray}
    d^D_1&=&\frac{i}{16},~d_2^D=-\frac{5}{1024},~d_3^D=-\frac{11i}{16384},~d_4^D=\frac{63}{524288},~d_5^D=\frac{527i}{20971520}\;,\nonumber\\
    d_6^D&=&-\frac{3129}{536870912},~d_7^D=-\frac{175045i}{120259084288},~d_8^D=\frac{422565}{1099511627776}\;,\nonumber\\
    d_9^D&=&\frac{1398251i}{13194139533312},~d_{10}^D=-\frac{266149}{8796093022208}\;.
\end{eqnarray}

%%%%%%%%%%%%%%%%%%%%%%%%%%%%%%%%%%%%%%%%%%%%%%
%%%%%%%%%%%%%%%%%%%%%%%%%%%%%%%%%%%%%%%%%%%%%%
%%%%%%%%%%%%%%%%%%%%%%%%%%%%%%%%%%%%%
\bibliography{refs}

\end{document}